\documentclass[showpacs,amsmath,amssymb,twocolumn,aps]{revtex4}

 \usepackage{graphicx}
\usepackage{epsfig}
 \usepackage{pstricks}
 \usepackage{pst-plot}
 \usepackage{psfrag}
 \usepackage{dcolumn}
 \usepackage{bm}

 \def\comment2#1{}
  \def\mn#1{}

\begin{document}

\title{Optical Phonon Lineshapes  and Transport 
in  Metallic Carbon Nanotubes under High Bias Voltage}

\author{J\"urgen Dietel}
\affiliation{Institut f\"ur Theoretische Physik,
Freie Universit\"at Berlin, Arnimallee 14, D-14195 Berlin, Germany}
\author{Hagen Kleinert}
\affiliation{Institut f\"ur Theoretische Physik,
Freie Universit\"at Berlin, Arnimallee 14, D-14195 Berlin, Germany}
\affiliation{ICRANeT, Piazzale della Repubblica 1, 10 -65122, Pescara, Italy}
\date{Received \today}

\begin{abstract}
We calculate the current-voltage characteristic  
of metallic nanotubes lying on a substrate
at high bias voltage showing that a bottleneck exists for 
short nanotubes in contrast to large ones. We attribute this 
to a redistribution of lower-lying acoustic phonons 
caused 
by phonon-phonon
scattering with hot optical phonons. 
The current-voltage characteristic
and the 
 electron and phonon distribution functions
are derived analytically, and serve to obtain 
in a self-contained way the 
frequency shift and 
line broadening of the zone-center optical 
phonons due to the electron-phonon coupling at high bias. We obtain 
a positive offset on the zero bias shift 
and no broadening of the optical phonon mode 
at very high voltages, in agreement with recent experiments.

\end{abstract}

\pacs{63.22.Gh, 78.30.Jw, 73.63.Fg, 73.50.Fq}

\maketitle

\section{Introduction}
Carbon nanotubes are 
one of the strongest and stiffest materials 
which can sustain very high currents before breaking. This 
electric property makes  
metallic nanotubes an interesting alternative  
to nanometer-sized metallic wires. Since 
nanotubes can behave like semiconductors,   
their possible use
in logic 
electronic circuits are promising. 
This has recently led  
to a number of experiments evaluating 
their current vs voltage characteristic at high bias voltage
\cite{Yao1, Perebeinos1, Park1, Javey1, Pop1, Bushmaker1,
 Oron-Carl1, Steiner1,
Sundqvist1}, with related
theoretical work in 
Ref.~\onlinecite{Lazzeri1, Lazzeri2, Auer1, Sundqvist2, Kuroda2, Kuroda1}.   

At low voltage, the current-voltage characteristic is mainly 
influenced by acoustic phonons and by
impurity scattering. At higher voltage, optical phonons 
become important. For metallic nanotubes on a 
substrate, the current vs voltage curve is increasing, 
in contrast to suspended nanotubes where the characteristic   
shows a negative differential conductivity at high bias 
\cite{Pop1}. 
     
We shall
review in Sect.~II the current-voltage characteristic 
of metallic nanotubes lying on a substrate. Following Refs. 
\onlinecite{Lazzeri2, Auer1} we use a Boltzmann approach for 
the electrons coupled to zone-center  and zone-boundary 
optical phonons. We take into account 
explicitly the dynamics of the phonons 
by a Boltzmann equation containing  an inelastic term 
to 
describe the decay of optical phonons into underlying 
acoustic phonons \cite{Lazzeri2, Auer1}. 
\comment2{das haben Refs. \onlinecite{Lazzeri2, Auer1}
doch auch getan, oder?}
We use first the so-called 
single-mode relaxation time 
approximation for the scattering term \cite{Srivastava1}.
This is characterized by 
a thermal phonon relaxation time $ \tau_{\rm op} $. 
For the  electron-phonon relaxation time 
$ \tau_{\rm ep} $
we use a numerically determined value 
 \cite{Lazzeri1, Pisanec1}, which 
reproduce very well the experimentally  determined lifetimes of optical 
phonons. This proceeding agrees 
with the numerical work of 
Refs.~\onlinecite{Lazzeri2, Auer1}
in which the current-voltage 
characteristic of short nanotubes with lengths smaller than 1 $ \mu $m 
is calculated. 
In contrast to this, Sundqvist {\it et al.} \cite{Sundqvist2} 
have in their calculation
$ \tau_{\rm ep} $-values 
which 
are around three times smaller than the experimental values. 
They determine the current-voltage characteristic of  nanotubes larger than 
1 $ \mu $m using the one-valley approximation
for the electrons. This
implies that the electrons are scattered 
only by one type of phonons, i.e. zone-center phonons,  between the bands 
within this valley. 
By using the experimentally determined thermal phonon relaxation 
life-time of $ \tau_{\rm op} \approx 1.1 \pm 0.2 $ps \cite{Song1,
  Kang1}, we reach
a good agreement
with  the experimentally determined 
current-voltage characteristics of large nanotubes.
This is in contrast with what happens in short nanotubes, 
which one has to use  
at least a five-times larger thermal phonon relaxation time to find a 
reasonable agreement with experiment. 

Due to the low dimensionality of a 
carbon nanotube system in which a fast initial decay of optical 
phonons is followed \comment2{so doch oder?}by a 
slow decay of only a small amount of secondary acoustic phonons 
\cite{Bonini1},
we expect a bottleneck in the relaxation path 
for the hot optical phonons generated by charge carrier scattering.
This idea was used in Ref.~\onlinecite{Bonini1} to explain the large 
discrepancy between the radial breathing mode lifetimes measured by  
Raman scattering experiments, and by tunneling experiments.  
Such a bottleneck leads of course to larger 
effective thermal relaxation times for the optical phonons. 
In Section III we shall  
describe this fact effectively by taking into account in the phonon 
Boltzmann  description the secondary acoustic phonons in a simple model. 
By using suitable secondary phonon relaxation times, we 
were able to reproduce 
the experimentally determined 
 current-voltage curve also for short tubes. Our result 
shows 
that  for long tubes the system does not exhibit a phonon bottleneck,
in contrast 
to short nanotubes. We explain this by the fact that at tube length 
smaller than $ 1 \mu $m, the thermal scattering lengths of 
many acoustic phonons 
reaches the systems size which then dynamically closes relaxation paths 
for the optical phonons.    

We find a similar effect in the interaction of phonons with the electron
system under bias voltage. It was shown in Refs. \onlinecite{Lazzeri2, Auer1}
for short nanotubes that one finds 
a large increase of the phonon distribution function 
at the boundaries of the tube. We observe an even worse
situation,
 that we do not find any numerical solution for the Boltzmann equation 
when setting the optical phonon velocities to zero. In contrast to this, 
we see for long tubes a phonon distribution function 
which is peaked in the center 
of the nanotube, in agreement with experiments for large suspended tubes
\cite{Deshpande1}. In order to understand this effect better, we 
solve in Section IV 
the system of Boltzmann equations for the charge carriers and the 
phonons analytically within certain approximations. We succeeded in 
reproducing
especially well the large-voltage small-length regime 
of the numerical determined current-voltage curves.
Our calculation shows that the reason for the increase of the 
temperature at the boundary of tube is again based on 
the fact that
using the phonon relaxation path for small tubes,
the electron phonon coupling part in the phonon Boltzmann 
equation creates effectively an additional phonon relaxation term with 
a negative sign.
This leads to the increase of the phonon temperatures at the boundaries of 
small tubes. 

In the analytical calculations of Section IV 
we determine the electron distribution function under high bias voltages. 
The appearance of this  distribution function is of course 
much different from the Fermi distribution function in thermodynamical 
equilibrium.  
The knowledge of this function opens up a number of possible 
applications. For example, in Section V, we calculate  
the level broadening and frequency shift
of the zone-center optical phonons 
mediated by the electron-phonon 
interaction under high bias voltage. We find a positive frequency  
offset on the zero bias shift for very large bias voltages, 
in contrast to the frequency shift 
mediated by the phonon-phonon scattering with phonons in thermal equilibrium.
For very large nanotubes we obtain also a negative frequency offset 
due to the electron-phonon interaction. 
The electron-phonon mediated 
zero-bias broadening of the zone-center optical mode 
vanishes at high voltages. The results for very high voltages 
are in agreement  with a recent 
experiment measuring the influence of the high bias on 
the phonon modes of carbon nanotubes lying on a substrate \cite{Oron-Carl1}.    
 
Summarizing, in Sect.~II we discuss the results of 
the coupled electron-phonon Boltzmann system in the relaxation 
time approximation numerically. In 
Section III the secondary 
acoustic phonons in the Boltzmann equation ia taken into account. 
We shall carry out in 
Section IV an analytical calculation of the current-voltage characteristic, 
the electron and the phonon distribution functions.  
These functions are used in Section V to calculate the level broadening 
 and frequency shift of the optical zone-center phonons  
mediated by the electron-phonon interaction under high bias voltage.

\section{Current-voltage characteristics of carbon nanotubes}
The method we use here to calculate 
the current-voltage characteristic of metallic nanotubes is based on 
the semi-classical Boltzmann equation. Within this method quantum 
interference corrections to the conductivity are not taken into account 
\cite{Datta1}. It was shown just recently through numerical 
calculations that these corrections to the conductivity 
are negligible above room temperature for single-walled 
carbon nanotubes without structural defects 
due to phonon scattering decoherence mechanisms \cite{Ishii1}. 
Other works using the semi-classical Boltzmann equation 
for electron or phonon transport not mentioned yet are found in  
Refs. \onlinecite{Pennington1, Vandecasteele1}.       

The energy levels of electrons in a nanotube consists of one-dimensional 
bands positioned in the graphene Brillouin zone around the $ {\bf K} $   and 
$  {\bf K'} $  points. For metallic nanotubes two energy bands 
corresponding to right (R) and left (L) moving electrons cross at these 
points. In the following, we assume here that the diameter 
$ D $ of the nanotube and the applied bias voltage $ U $ is so small 
that we can neglect electron excitations to higher bands. 
For example, this is valid for a nanotube with diameter 
$ D\approx 2$nm when we apply a bias voltage of less similar to   
$ U \lesssim 2 $V. 
The electron distribution functions for a nanotube under bias voltage 
around the $ {\bf K} $ and $ {\bf K'} $ points are equal  
which we denote by $ f_{L/R}(k,x ,t) $.
At larger voltages only the optical phonons are relevant
as a source of electron-hopping between the bands.  
The hopping between one band at 
$ {\bf K} $ and the other at the $ {\bf K'}$-point 
 are mediated by zone-boundary 
optical phonons where only Kekul\'{e} type of lattice distortions
couple to the electronic system \cite{Suzuura1}.
We denote in the following the corresponding 
phonon distribution function by $n^{\rm K}(k,x,t) $.  
The hopping between bands at the same  
$ {\bf K} $ or $ {\bf K'} $ points  are mediated by longitudinal zone-center  
optical phonons with phonon distribution $n^\Gamma(k,x,t) $ \cite{Lazzeri1}. 
The time evolution of the electrons are governed by the 
semi-classical Boltzmann-equation   
\begin{equation}
\left(\partial_t \mp v_F \partial_x +\frac{e {\rm E}}{\hbar} 
  \partial_k\right) 
f_{L/R}= \left[\partial_t f_{L/R}\right]_c        \label{10}
\end{equation} 
where the collision term    
$ [\partial_t f_{L/R}]_c \approx [\partial_t f_{L/R}]_e+ 
+[\partial_t f_{L/R}]_{\rm fs}+ 
[\partial_t f_{L/R}]_{\rm bs} $ consists of an elastic 
scattering term 
$ [\partial_t f_{L}]_e = v_F/\l_e [f_R(k)-f_L(k)] $  due to acoustic phonon 
scattering (in the quasi-elastic limit) 
and impurity scattering. $ l_e $ is the 
elastic scattering mean free path. 
We assume here $ l_e = 1600\, $nm \cite{Lazzeri2, Park1}.  
The electron velocity $ v_F $ is given by $ v_F=8.4 \times 10^7 \,$cm/s and 
$ e $ ($ e>0$) is the electronic charge.

$ [\partial_t f_{L/R}]_{\rm fs} $ is a forward scattering 
term which should have a minor effect especially for higher applied voltages
since it does not change the propagation direction. 
The time evolution of optical phonons is  given by  
\begin{equation}
\left[\partial_t +v^\nu_{\rm op}(k) \partial_x\right] 
n^{\nu}= \left[\partial_t n^\nu\right]_c +
\left[\partial_t n^\nu\right]_{\rm osc} \,.         \label{20}
\end{equation}
where $ \nu=\Gamma,{\rm K} $ denotes zone-center or zone-boundary phonons.
We use here $ v^\Gamma_{\rm op}= {\rm sign}(k) \, 2.9 \times 10^5 \, $cm/s and 
$ v^{\rm K}_{\rm op}= {\rm sign}(k)\, 7.2 \times 10^5 \, $cm/s 
\cite{Lazzeri2,Pisanec1}.
The term $ \left[\partial_t n^\nu\right]_c $ is 
due to phonon-electron  scattering, while the term  
$ \left[\partial_t n^\nu\right]_{\rm osc} $ represents thermal phonon 
relaxation. Note that the coupled electron-phonon system is not heated up 
by applying large voltages on the nanotube due to this term, which accounts  
effectively for 
the scattering of optical phonons into underlying 
(acoustical) phonons.      

Scattering of phonons with electrons leads to two scattering contributions
in the electronic Boltzmann equation (\ref{10}) as well as in the phononic 
Boltzmann equation (\ref{20}). When restricting on the backward scattering 
contributions we obtain for the electronic scattering term   

\begin{align}
&  
\left[\partial_t f_{L}\right]_{\rm bs}
=\sum_\nu \frac{1}{\tau^\nu_{\rm ep}}\times  \label{30} \\  
& 
 \bigg\{[n^\nu(k^+,x)+1]  f_R(k_R(\epsilon^{+}))
[1-f_L(k_L(\epsilon))]\,       \nonumber \\
& -n^\nu(k^+,x)[1-f_R(k_R(\epsilon^{+})) ]  
f_L(k_L(\epsilon)) \,             \nonumber \\    
& +
n^\nu(-k^-,x)  f_R(k_R(\epsilon^{-})) 
[1-f_L(k_L(\epsilon))] \,    \nonumber \\
& -[1+n^\nu(-k^-,x)][1-f_R(k_R(\epsilon^{-}))]
f_L(k_L(\epsilon))    \bigg\}     \nonumber 
\end{align} 
with $ k^\pm=k_R(\epsilon^{\pm })  
-k_L(\epsilon) $ and $ \epsilon^{\pm}= \epsilon \pm  \hbar \omega^\nu $.
The corresponding phononic scattering term results in    
\begin{align}
&  
\left[\partial_t n^{\nu}\right]_c
=\sum_\nu \frac{s^\nu}{\tau^\nu_{ep}}    
\times \bigg( [n^\nu(k,x)+1]\times
      \nonumber \\
& \times\left\{f_R(k_R^+)  
[1-f_L(k_L^-)]
+f_L(-k_L^-)
[1-f_R(-k_R^+)] \right\}\!   \nonumber   \\  
&   
 \hspace{26ex}- \! n^\nu(k,x) \times \nonumber   \\  
&   
\!\times \left\{f_L(k_L^-)
[1-f_R(k_R^+)]   +f_R(-k_R^+)
[1-f_L(-k_L^-)]\right\} \bigg), \label{40}
\end{align}  
where $ k_{R/L}^{\pm}=k_{R/L}(\epsilon(k/2)\pm\hbar \omega^\nu/2) $.
The number $ s^\nu $ is given by $ s^{\rm K}=1 $ for zone-boundary phonons 
and $ s^\Gamma=2 $ for zone-center ones \cite{Lazzeri2}. 
In order to derive these numbers one has to take into account that 
momentum phase space of the phonons is twice as large as 
the phase space of the electrons. Further one has to consider 
the fact that the electron jumps 
from the $ {\bf K} $ to the $ {\bf K'}$-band are mediated by 
$ {\bf K}$-phonons but the reverse jumps by $ {\bf K}'$-phonons. 
On the other hand  jumps of electrons within the same valley are mediated 
by the same $ {\bf \Gamma} $-phonons. 
Finally, we mention here that we used the boundary conditions \cite{Lazzeri2}  
\begin{align} 
& f_L(k_L(\epsilon),L) = f_R(k_R(\epsilon),0)= n_F(\epsilon) \,,  \label{42} \\
& n^\nu(k>0, 0) = n^\nu(k<0, L)=n^{\rm op}_B  \,.                       \label{45} 
\end{align} 
where $ n_F(\epsilon) $ is the Fermi function for a metallic nanotube 
with zero gate voltage at room temperature, i.e. 
$ n_F(\epsilon)=1/(1+e^{\epsilon/k_B T}) $, and   
$ n^{\rm op}_B $ is the Boltzmann factor for optical phonons 
at room temperature given by $ n^{\rm op}_B \approx 0.0014$.
We use optical phonon frequencies $ \hbar \omega^{\rm K}= 161 \, $meV and 
$ \hbar \omega^\Gamma= 196 \, $meV. 
The electron-phonon scattering times for zone-center and zone-boundary 
optical phonons are given by 
$ \tau_{\rm ep}^\Gamma = 538 \, $fs and $ \tau_{\rm ep}^{\rm K} = 219 \, $fs 
\cite{Lazzeri1} where we assume tube diameters of around 2.0nm typical 
in existing current-voltage experiments in the literature.

Our method to solve (\ref{10}) and (\ref{20})
is based on the numerical time integration by the standard splitting method 
\cite{Aristov1}.
We discretize the differential equations in momentum 
and position space \cite{Rem1}. To integrate the collisionless 
free electron and phonon equations in some time step, we use the 
exact solution of the equations in the case of the electron motion. 
This means that the time step value are fixed by the space grid. 
The free phonon motion in one time-step is given by the up integration 
of the collisionless discrete version of (\ref{20}) on 
the space grid.       

In this section, we use a standard 
single-mode relaxation time approximation for the optical 
phonon scattering term $ \left[\partial_t n^\nu\right]_{\rm osc} $ given by 
\begin{equation} 
 \left[\partial_t n^\nu\right]_{\rm osc}= -\frac{1}{\tau_{\rm op}} 
(n^\nu-n^{\rm op}_B)\,. 
\label{50}
\end{equation}     
Note that this approximation is only valid 
for the system lying on a substrate. 
For the suspended nanotube system one has to take into account 
explictly the heat transfer by acoustic phonons to the leads \cite{Pop1}. 
We assume in our calculation that the thermal relaxation time 
$\tau_{\rm op} $ is similar for zone-center and zone-boundary optical 
phonons. This approximation is justified for graphene in Ref. 
\onlinecite{Bonini2} 
where it is shown that the relaxation times of both phonon types 
are almost equal for 
acoustic phonon temperatures a little 
higher than the room temperature. 
We do not expect a difference for carbon nanotubes. 
These temperatures are immediately reached at the high voltage experiments 
we are interested in \cite{Bushmaker1, Oron-Carl1, Deshpande1, Steiner1}.

First we calculate the current-voltage characteristic for 
nanotubes of length larger than $ 1 \mu $m in the vicinity of  
$ \tau_{\rm op} =1.1 $ps. This value is chosen  
since Song {\it et al.} \cite{Song1} 
determined experimentally $ \tau_{\rm op} =1.1 \pm 0.2 $ps  
in agreement to the experiment of 
Kang {\it et al.} in Ref.~\onlinecite{Kang1}. 
     
In the upper panel in Fig.~1 the current-voltage 
characteristic determined with help of Eqs.  
(\ref{10})-(\ref{50}) is shown for nanotubes
at bias voltage $ U=E L =1 V $ as a function of their length for various 
relaxation times $ \tau_{\rm op} $ in the vicinity of the experimentally 
determined relaxation time.  
The solid curve in the Figure 
is given by the experiment carried out by Sundqvist 
{\it et al.} in Ref.~\onlinecite{Sundqvist1}. 
In this  experiment the nanotube length was effectively varried by changing   
the distance between the electrodes where the bias voltage is applied.  

We get the best agreement within 
the experimental uncertainties for the quantity $ \tau_{\rm op} =1.1\pm 0.2 $ps
at value $ \tau_{\rm op}=0.9 $ps. 
In the lower panel in Fig.~1, we calculate  the current-voltage characteristic of nanotubes
for $ \tau_{\rm op}=0.9 $ps as a function of the tube length for various
bias voltages. Fig.~2 shows the average phonon density for a nanotube
at length $ L=3000$nm, $ U=1V $ and $ \tau_{\rm op}=0.9 $.    
This density is determined by 
\begin{equation} 
\overline{n}^\nu(x)=\frac{1}{2 e E L} 
 \int\limits_{e E (x-L)}^{e E x}
 \! \!  d \epsilon \, [n^\nu(2 k_L(\epsilon),x)+
n^\nu(2 k_R(\epsilon),x)]  \label{55} 
\end{equation} 
The factor two in the denominator is necessary due to the fact 
that we average over the right and left moving  electron 
bands. We obtain phonon densities which are peaked in the 
center of the nanotubes. This behaviour is in  
accordance to experiments \cite{Deshpande1} for suspended nanotubes. 
We show also in this figure the space 
averaged phonon densities $ \overline{n}^\nu= 
\int dx \overline{n}^\nu(x)/L $ as a function of the tube 
length for various bias voltages. 

One reason for the small difference in the current-voltage 
characteristic between experiment and theory in the upper panel  of 
Fig.~1 at large nanotube lengths is due to the fact that 
the diameter $ D $ of the tube 
in the experiment Ref.~ \onlinecite{Sundqvist1} is in fact 
a little larger than $ 2 $nm. Although the diameter 
was not measured explicitly in Ref.~ \onlinecite{Sundqvist1} 
one can estimate it by the fact that Sundqvist {\it et al.} 
measured approximately half of  
the differential resistivity for short 
distances between the elctrodes in comparison to the value in 
Refs.~\cite{Park1, Javey1}. In these experiments 
the current-voltage characteristic of short nanotubes 
with a measured diameter $ D \approx 2 $nm 
was recorded. By using an analytical theory for the current-voltage 
characteristic which will be derived in Sect.~IV and further that the 
electron-phonon scattering time $ \tau^{\nu}_{\rm ep} $ 
is proportional to the diameter $ D $ of the nanotube \cite{Lazzeri1} 
we obtain that $ D\approx \sqrt{2} \times 2 {\rm nm} \approx 2.8 $nm.
We show in the upper panel in Fig.~1 by the dotted curve 
the theoretically calculated current-voltage 
characteristic for a $ 2.8 $nm nanotube, i.e.  
$ \tau^\nu _{\rm ep} $ is now a factor $ \sqrt{2} $ larger 
than for the $ 2.0$nm nanotube already used before, 
and $ \tau_{\rm op}=0.9 $ps. 
 
In contrast to the small undershooting of the theoretically determined 
current-voltage curve in comparison to the experimental curve for large 
lenths in Fig.~1, we obtain for 
small nanotube lengths an overshooting of the curve. 
The reason for this different behavior between 
large and small nanotube lengths will be discussed in the following.        
 
Next, we determine the current-voltage characteristic 
of short nanotubes. In Fig.~3 we show the current-voltage 
characteristic for a nanotube of length $ L=300 $nm as a function of the 
bias voltage $ U $ for various phonon relaxation times $ \tau_{\rm op}$.  
The solid curve is given by the experiment \cite{Javey1}.
We obtain the best agreement between experiment and theory    
for the differential conductivity $ dU/dI \approx 220 k \Omega $
at $  \tau_{\rm op} \approx 9.1 $ps.   
It is astonishing that this optical phonon relaxation time 
 is  much larger than the 
experimentally determined optical phonon relaxation time 
$ \tau_{\rm op} = 1.1 \pm 0.2 $ps.
The reason for this discrepancy will be 
discussed in the next section.  
Note that we obtain in Fig.~3 in the low voltage regime a better 
agreement between the 
theoretically and experimentally determined curves by using smaller elastic 
scattering lengths $ l_e< 1600$nm.  

\begin{figure}
\begin{center}
 \includegraphics[height=8cm,width=8.5cm]{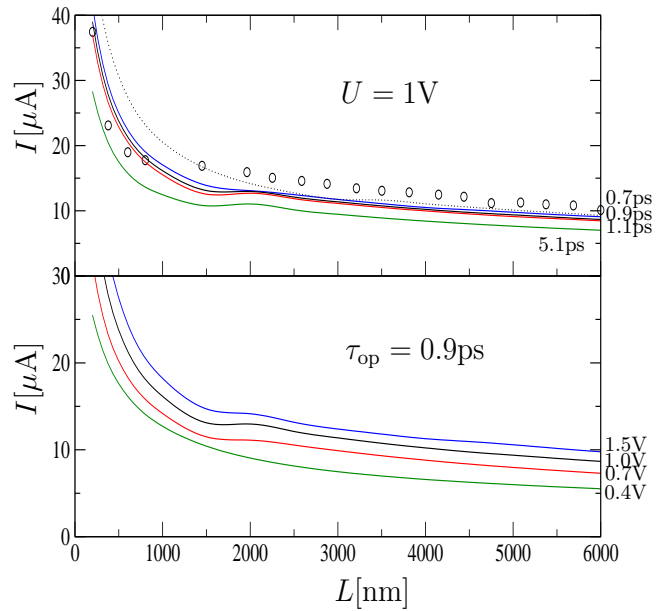}
\caption{(Color online) 
Upper panel shows the current-voltage characteristic of 
long nanotubes with diameter $ D=2.0$nm for bias voltage $ U=1$V 
calculated by the help of (\ref{10})-(\ref{50}) 
for various thermal relaxation times $ \tau_{\rm op} $ (solid curves)
and $ l_e=1600 $nm. The dotted curve shows 
the current-voltage characteristic of a $ D=2.8$nm nanotube, i.e. it uses 
$\sqrt{2} 
\tau^\nu_{\rm ep} $ as the electron-phonon scattering times, and 
$ \tau_{\rm op} = 0.9 $ps. 
The (black) circles are given by the experiment \cite{Sundqvist1}.
The lower panel shows the  
current-voltage characteristic of long nanotubes with diameter $ D=2.0 $nm 
for various bias voltages, $ \tau_{\rm op}=0.9$ps and $ l_e=1600 $nm.}   
\end{center}
\end{figure} 

\begin{figure}
\begin{center}
 \includegraphics[height=7.5cm,width=8.5cm]{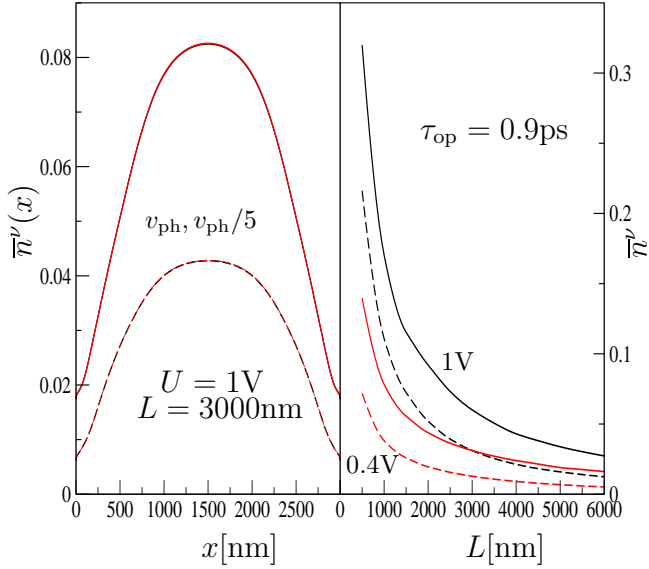}
\caption{(Color online) 
Left panel shows the phonon-density distribution functions 
$ \overline{n}^K(x) $ 
(solid curves) and $ \overline{n}^\Gamma(x) $ (dashed curves) defined in (\ref{55})
for $ U=1$V, $ \tau_{\rm op}=0.9 $ps and $ L=3000$nm. 
The black curve is calculated by 
using the former defined  $ v^\Gamma_{\rm op}= {\rm sign}(k) \, 2.9 \times 10^5 \, $cm/s and 
$ v^{\rm K}_{\rm op}= {\rm sign}(k)\, 7.2 \times 10^5 \, $cm/s 
\cite{Lazzeri2,Pisanec1}. The red curves uses $ v^\Gamma_{\rm op}/5 $ and 
$ v^{\rm K}_{\rm op}/5 $ as phonon velocities. The curves lie practical
on top of each other. The right panel shows position averaged phonon 
distribution functions $ \overline{n}^{\rm K}$ (solid curves) and 
$ \overline{n}^\Gamma$ (dashed curves) for $ U=0.4 $V (lower red curves) and 
$ U=1V $ (upper black curves) for $ \tau_{\rm op}=0.9 $ps.       
}   
\end{center}
\end{figure}

\begin{figure}
\begin{center}
 \includegraphics[height=6cm,width=8.5cm]{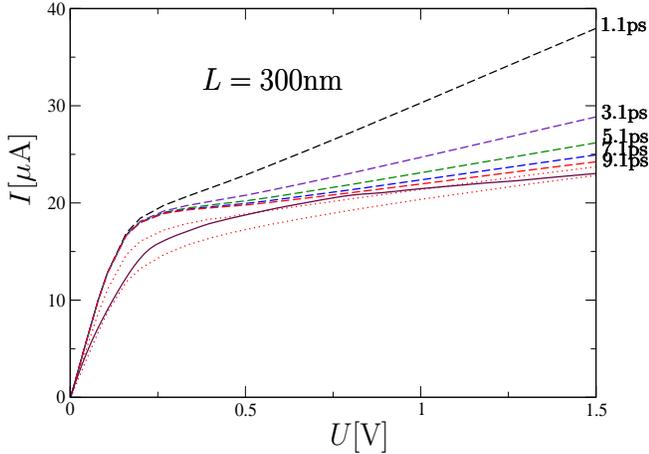}
\caption{(Color online) Current-voltage characteristic of 
a $ L=300$nm nanotube calculated by the help of (\ref{10})-(\ref{50}) 
for various thermal relaxation times 
$ \tau_{\rm op} $ (dashed curves) and $ l_e=1600 $nm. 
The (black) solid curve is given by the experiment \cite{Javey1}. The (red) 
dotted curves are calculated for $ \tau_{\rm op}=9.1$ps, 
$ l_e=800 $nm (upper curve) and $ l_e=400 $nm (lower curve).}   
\end{center}
\end{figure}

\section{Second generation phonons}
From Fig.~3 we see that a satisfactory agreement between the 
experimentally and numerically determined  
current-voltage characteristic is only reached for 
$ \tau_{\rm op} \gg 1.1$ps. On the other hand, recent 
phonon lifetime experiments on carbon nanotubes show that 
$ \tau_{\rm op} \approx 1.1$ps for zone-center phonons 
\cite{Song1, Kang1}. These measured 
phonon lifetimes are governed by the decay of 
zone-center phonons to two lower energetic second-generation 
phonons where the number 
of these decay channels should be rather small for one-dimensional 
nanotube systems in contrast to higher dimensional systems like 
graphene or graphite \cite{Bonini1}. The second-generation phonons 
are typically acoustic ones which then again 
scatter in two acoustic phonons 
with even lower energy and  
longer wave-length where this lifetime   
is much longer than of the primary 
optical phonons. The reason for the longer lifetime
comes from the fact that the three phonon matrix element vanishes 
in the long-wavelength limit and further that the phase space for 
phonon decay  is smaller for 
lower phonon energies  due to energy conservation. The long lifetime
of secondary phonons and the small amount of possible decay channels could 
lead to a bottleneck in the decay process. 
This means that 
a significant amount of secondary phonons are assembled in the decay of  
hot-phonons generated by 
charge carriers through the electron-phonon interaction. 
When this non-equilibrium amount of secondary phonons is similar 
to the number of equilibrium phonons following the Bose-Einstein 
distribution the single-mode relaxation time method 
leading to the appearance  of the scattering expression (\ref{50})  
is no longer valid. In Ref.~\onlinecite{Bonini1} it was argued 
that this fact is responsible for the considerable difference 
in the lifetime measurements of the radial breathing mode  
by using either Raman-scattering experiments or electron tunnel experiments. 
The decay channel can then be described by the following Boltzmann equations 
\cite{Jursenas1} when neglecting the phonon velocities 
$ v^\nu_{\rm op} \approx 0 $  
on the left-hand side of~(\ref{20}): 
\begin{align} 
& \partial_t n^\nu= \frac{1}{\tau_{\rm op}}\left[ -n^\nu (1+n_{\rm ac})^2
+ (1+n^\nu) n^2_{\rm ac}\right] +\left[\partial_t n^\nu\right]_c \label{1010} 
\,,  \\
& \partial_t n_{\rm ac} = 
\frac{1}{p \tau_{\rm op}}\left[ n^\nu (1+n_{\rm ac})^2
- (1+n^\nu) n^2_{\rm ac}\right]           \nonumber \\
& \qquad \qquad - \frac{1}{\tau_{\rm ac}}(n_{\rm ac}-n^{\rm ac}_B)         \,.        
\label{1020} 
\end{align}        
Here the first term in the brackets in (\ref{1010}) 
describes the scattering of the optical 
phonons with distribution function $ n^\nu $ into two secondary phonons 
with distribution function $ n_{\rm ac} $.  
For simplicity we assumed that the secondary phonons 
follow all the same distribution function.

The second term in the brackets in (\ref{1010}) describes the reverse process. 
The second equation (\ref{1020}) describes the dynamics of the secondary 
phonons.  
Here $ p $ denotes the number of decay channels. For 
one-dimensional solids this number is generally small \cite{Bonini1}. 
For simplicity we further assumed in (\ref{1020}) 
that the secondary phonons are coupled 
to a heat bath where the relaxation with this bath happens in time 
$ \tau_{\rm ac} \gg \tau_{\rm op} $. 
The quantity
$ n^{\rm ac}_B $ is the Bose-factor for the secondary phonons 
which we assume to be half of the frequency of the optical primary 
phonons,
leading to $ n^{\rm ac}_B \approx 0.03 $ at room temperature.
Note that this choice is consistent with the fact that we choose 
uniform secondary phonon distributions.   

In the stationary case we have $ \partial_t n_{\rm ac}=0 $. 
Then we can solve the second equation (\ref{1020}) for  
$ n_{\rm ac} $ and insert the result into the first equation 
which leads to an effective optical phonon scattering term    
 \begin{align} 
&  \left[\partial_t n^\nu\right]_{\rm osc}= -\frac{1}{\tau_{\rm op}} \left[(n^\nu-n^{\rm ac}_B)\frac{p \tau_{\rm op}}{\tau_{\rm ac}}- \frac{1}{2}
\left(\frac{p \tau_{\rm op}}{\tau_{\rm ac}}\right)^2  \right. 
 \label{1030}   \\
& \left. + \frac{1}{2} 
\frac{p \tau_{\rm op}}{\tau_{\rm ac}} 
\sqrt{\left(\frac{p \tau_{\rm op}}{\tau_{\rm ac}}- 2n^\nu\right)^2 +
4 \left(n^\nu+\frac{p \tau_{\rm op}}{\tau_{\rm ac}} n^{\rm ac}_B   \right)} 
\right]    \nonumber 
\end{align}
with limits
\comment2{Unterschied beiden Formeln erklaeren}
\begin{align}
&  
\lim_{p \tau_{\rm op}/\tau_{\rm ac} \to \infty } 
\! \! \! \! \! \! \! \! \! 
\left[\partial_t n^\nu\right]_{\rm osc} \rightarrow   - \frac{1}{\tau_{\rm op}}
\left[n^\nu (1+2 n^{\rm ac}_B)-(n^{\rm ac}_B)^2\right]  \,,  \label{1040} \\
 & 
 \lim_{p \tau_{\rm op}/\tau_{\rm ac} \to 0} 
\! \! \! \! \! \! \! \! \! 
 \left[\partial_t n^\nu\right]_{\rm osc}  \rightarrow   - \frac{1}{\tau_{\rm op}}
 \frac{p \tau_{\rm op}}{\tau_{\rm ac}} \left[\! \! \left(1\!+\!\sqrt{1\!+\!\frac{1}{n^\nu}} \! \right)
 n^\nu -n^{\rm ac}_B \! \right] \! . \label{1050}  
\end{align} 
As is seen from (\ref{1040}) in the case of no existent bottleneck, 
i.e. large 
$ p \tau_{\rm op}/\tau_{\rm ac} $ and small 
Boltzmann factors $ n^{\rm op}_B $, $ n^{\rm ac}_B $ valid in our case, 
we obtain the 
standard single-mode relaxation time approximation (\ref{50}) for the optical 
phonon scattering term.   
 
\begin{figure}
\begin{center}
 \includegraphics[height=9cm,width=8.2cm]{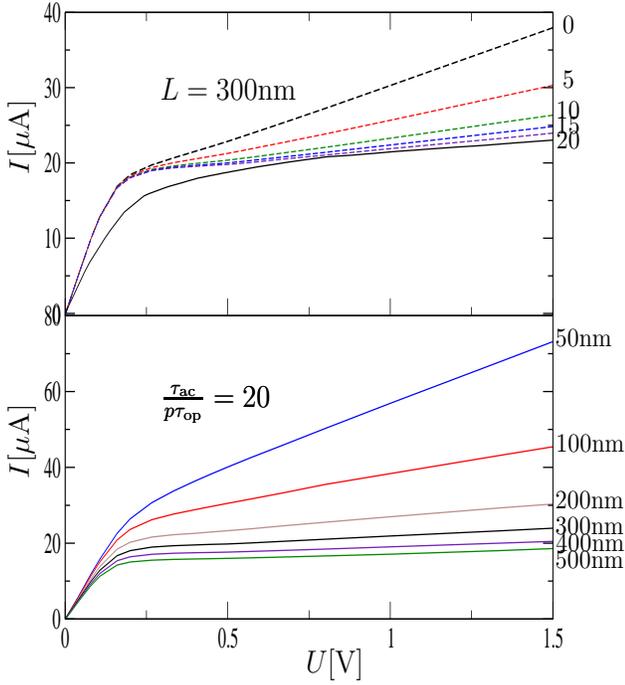}
\vspace*{0.2cm} 
\caption{(Color online) 
Upper panel shows the current-voltage characteristic of 
a $ L=300$nm nanotube calculated by the help of (\ref{10})-(\ref{40}) 
and (\ref{1030}) for various parameters  
$ \tau_{\rm ac}/p \tau_{\rm op}  $ 
(dashed curves) and  $ \tau_{\rm op}=1.1$ps, $ l_e=1600 $nm. 
The (black) solid curve is given by the experiment \cite{Javey1}.
The lower panel shows the  
current-voltage characteristic of various nanotubes
with different lengths for $ \tau_{\rm op}=1.1$ps  
and $ \tau_{\rm ac}/p \tau_{\rm op} =20 $, $ l_e=1600 $nm.}   
\end{center}
\end{figure} 

\begin{figure}
\begin{center}
 \includegraphics[height=8cm,width=8.2cm]{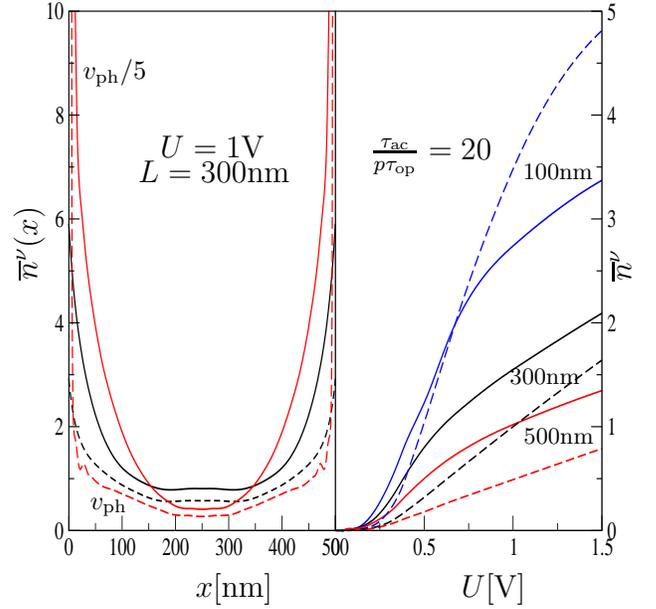}
\vspace*{0.2cm} 
\caption{(Color online) 
Left panel shows the phonon-density distribution functions 
$ \overline{n}^K(x) $ 
(solid curves) and $ \overline{n}^\Gamma(x) $ (dashed curves) defined 
in (\ref{55})
for $ U=1$V and  $ \tau_{\rm ac}/p \tau_{\rm op} =20 $,  
$ \tau_{\rm op}=1.1 $ps and $ L=300$nm. 
The black curve is calculated by 
using the former defined phonon velocities 
$ v^\Gamma_{\rm op}= {\rm sign}(k) \, 2.9 \times 10^5 \, $cm/s and 
$ v^{\rm K}_{\rm op}= {\rm sign}(k)\, 7.2 \times 10^5 \, $cm/s 
\cite{Lazzeri2,Pisanec1}. The red curves uses $ v^\Gamma_{\rm op}/5 $ and 
$ v^\Gamma_{\rm op}/5 $ as phonon velocities. 
The right panel shows position averaged phonon 
distribution functions $ \overline{n}^\Gamma$ (dashed curves) and 
$ \overline{n}^K$ (solid curves) 
for $ U=1 $V, $ \tau_{\rm ac}/p \tau_{\rm op} =20 $ 
and $ L=100 $nm (blue curves), $ 300 $nm (black curves) and 
$500 $nm (red curves). 
}   
\end{center}
\end{figure}

In the following, we carry out the numerical calculation by using 
(\ref{10})-(\ref{40}) with the optical phonon scattering term 
(\ref{1030}) substituting (\ref{50}). From (\ref{1030}) we obtain that the 
optical scattering term depends via  
$ \tau_{\rm ac}/ p \tau_{\rm op}  $ on the acoustic scattering length. 
We show in the upper panel of Fig.~4 the 
current voltage characteristic for various 
parameters $\tau_{\rm ac}/ p \tau_{\rm op} $,  
$ \tau_{\rm op} =1.1$ps and nanotubes of length 
$ L\approx 300$nm. This figure should be compared to  Fig.~3 in the case 
of the single-mode relaxation time approximation which uses (\ref{50})
for the optical phonon scattering term. We obtain a similar behavior of
the current-voltage characteristic curves in both approximations. 
The reason is seen from expression (\ref{1050}).
We obtain from this expression and the fact that at room temperature 
$ n^{\rm ac}_B, n^{\rm op}_B \ll 1$ as well as $ n^\nu \gg 1 $ 
for high bias voltages, that for small 
$ p \tau_{\rm op}/ \tau_{\rm ac} $ the effective 
optical phonon scattering term 
$ \left[\partial_t n^\nu\right]_{\rm osc} $ 
has still the standard single mode relaxation form (\ref{50}). 
The effective relaxation time $ \tau_{\rm op} $ 
is then changed to $ \tau_{\rm op} \to \tau_{\rm ac}/2 p $ which 
respects the fact that the relaxation of the optical phonons 
are effectively relaxed on $ 2 p $ channels of relaxation time
$ \tau_{\rm ac} $.  

In the lower panel of Fig.~4 we show the current-voltage 
characteristic for 
various nanotube lengths by using (\ref{1030}) as 
the optical phonon scattering term with $\tau_{\rm ac}/ p \tau_{\rm op} =20 $ 
and $ \tau_{\rm op}=1.1$ ps. 
Fig.~5 shows the energy-averaged phonon distribution function (\ref{55}) 
$ \overline{n}^\nu(x) $ (left panel) and the 
energy and position-averaged phonon distribution 
function $ \overline{n}^\nu $ (right panel) for various nanotube 
lengths. We obtain that 
in contrast to the case of large nanotubes in Fig.~2, the short 
nanotubes show an increasing phonon 
density at the boundary of the nanotube 
\cite{Lazzeri2, Auer1}. We obtain from the left panel in Fig.~5 the even worse 
fact of a diverging current-voltage characteristic for 
phonon velocities $ v^\nu_{\rm op} \to 0 $. We shall understand this unusual 
behaviour better in the next section where we show an analytic solution 
of the Boltzmann system (\ref{10})-(\ref{50}).

Summarizing, by taking into account a possible bottleneck in the 
optical phonon relaxation path we obtain for an optical phonon 
relaxation time $ \tau_{\rm op} = 1.1 $ps for large nanotubes an effective 
scattering parameter $\tau_{\rm ac}/ p \tau_{\rm op} =0 $ as the best fitting 
parameter to the experimental curves, i.e. no phonon bottleneck is seen
in this case. On the other hand for small nanotubes and 
$ \tau_{\rm op} = 1.1 $ps we obtain an effective 
scattering parameter of $\tau_{\rm ac}/ p \tau_{\rm op} =20 $.
The reason for this discrepancy between large and small nanotubes 
lies in the fact that we have neglected the velocities of the 
secondary phonons in the Boltzmann equation (\ref{1020}). This leads
in the case of the decay of zone-center optical phonons to  
additional terms of the form $ \pm v_{\rm ac} \partial_x n^\pm_{\rm ac} $ 
in the left hand side of (\ref{1020}) 
where $ n^\pm_{ac} $ stands for the left and right moving phonon 
in a scattering pair. This means that we remove the restriction 
in the right hand sides 
of (\ref{1010}) and (\ref{1020}) that the scattering pairs have all the 
same distribution function. To see when the 
$ \pm v_{\rm ac} \partial_x n^\pm_{ac} $ term becomes relevant in the Boltzmann equation we further have to take into account the number of decay channels
$ p $ for nanotubes. In the following we restrict ourselves to the relaxation 
of the zone-center phonons. The argument for the zone boundary phonons 
works similar.  

It was shown in Refs.~\onlinecite{Bonini1, Bonini2}  
that in the case of graphene the zone-center optical phonons scatter into 
three sorts of different pairs of phonons lying on rings in the Brioullin 
zone around the $ {\bf \Gamma} $  points where scattering 
into the longitudinal acoustic 
sector is in fact the most dominant. In order to estimate from this fact 
the number of pairs for a nanotube with diameter of around $ 2 $nm 
we use in the following the zone 
folding approximation method. By using as an approximation that in the 
case of graphene the decay rings lie in the mids between the 
$ {\bf \Gamma} $ and $ {\bf K} $ point  
(best fulfilled for the longitudinal mode \cite{Bonini1, Bonini2})
we obtain as an estimate for the number of decay pairs $ p \lesssim 20 $ 
in a $ 2 $nm nanotube. In order to see no bottleneck 
in the relaxation process for large 
nanotubes we have $ \tau_{\rm ac} \lesssim p \tau_{\rm op} $ meaning 
that $ \tau_{\rm ac} \lesssim 22 $ps. 
On the other hand for a nanotube 
of diameter $ 2$nm one obtains for the lowest lying phonon modes
in a simple model  relaxation times which are larger than around 
$ \tau_{\rm ac} \gtrsim 20 $ps 
\cite{Hepplestone1}. This leads us to the estimate  
$ \tau_{\rm ac} \sim 20 $ps for the effective acoustic relaxation time.  
With $ v_{\rm ac} \sim  21$km/s 
(we choose the maximum velocity value for acoustic 
phonons in graphite \cite{Saito1, Mahan1, Zhang1})
we obtain that the acoustic phonon velocity terms 
$ \pm v_{\rm ac} \partial_x n^\pm_{\rm ac} \sim 
\pm v_{\rm ac} n^\pm_{\rm ac}/L $ in the phonon Boltzmann 
equation becomes relevant for $ L \lesssim 420 $nm. 
This is only a very rough approximation for this length. 

At this length we find that for one participant of the scattered 
acoustic phonon pairs this additional relaxation term is not 
relevant since the relaxation path is already open.  
For the other participant 
this term leads to a closing of the relaxation path which we 
saw in our numerics as an increase of the effective relaxation parameter 
$ \tau_{\rm ac}/p \tau_{\rm op} $.  
  
\section{Analytic  calculation}

Due to the similarity of the phonon frequency of zone boundary phonons 
$ \omega^{\rm K} $ and zone-center phonons $ \omega^\Gamma $ and  
since the electron-phonon 
coupling constants $ s^{\rm K}/\tau^{\rm K}_{\rm ep} $ and 
$ s^\Gamma/\tau^\Gamma_{\rm ep} $ are similar we use in the 
following the simplification that the nanotube system interacts 
with only one sort of phonons 
with frequencies $ \omega $. The effective  electron-phonon scattering 
parameter $ \tau_{\rm ep} $ in the electronic sector, the 
electron-phonon scattering parameter $ s^p/\tau_{\rm ep} $  
in the phononic sector and phonon velocities $ v_{\rm ph} $ are chosen in 
the following way       
\begin{eqnarray} 
\hbar \omega  & =  & \hbar  
\left(\frac{\omega^{\rm K}}{\tau^{\rm K}_{\rm ep}}+\frac{\omega^\Gamma}
{\tau^\Gamma_{\rm ep}}\right)\bigg/
 \left(\frac{1}{\tau^{\rm K}_{\rm ep}}+\frac{1}{\tau^\Gamma_{\rm ep}}\right)  
= 170 {\rm meV} \,, \nonumber \\  
\frac{1}{\tau_{\rm ep}} & =
&\left( \frac{1}{\tau^{\rm K}_{\rm ep}}+ \frac{1}{\tau^\Gamma_{\rm op}}\right)=
\frac{1}{155{\rm fs}}\,, \nonumber \\
\frac{s^p}{\tau_{\rm ep}} & =
&\left( \frac{s^{\rm K}}{(\tau^{\rm K}_{\rm ep})^2}    + \frac{s^\Gamma}{(\tau^\Gamma_{\rm op})^2}\right)\bigg/
 \left(\frac{1}{\tau^{\rm K}_{\rm ep}}+\frac{1}{\tau^\Gamma_{\rm ep}}\right) =
\frac{1}{231{\rm fs}} \,, \nonumber \\
v_{\rm ph}& = &  
\frac{v^{\rm K}_{\rm ph}}{\tau^{\rm K}_{\rm ep}}+\frac{v^\Gamma_{\rm ph}}
{\tau^\Gamma_{\rm ep}}\bigg/\! \! 
 \left(\! \frac{1}{\tau^{\rm K}_{\rm ep}} \! + \! 
\frac{1}{\tau^\Gamma_{\rm ep}}\! \right)           \nonumber \\
& = & \rm{sgn}(k) \, 5.95 \times 10^5 \frac{{\rm cm}}{s} \,.  \label{68} 
\end{eqnarray}
In the following discussion, we use the abbreviation 
$ l^r_{\rm sc} \equiv v_F \tau_{\rm ep} /(2 \overline{n} +1) $ 
for the reduced effective scattering length and 
$ l_{\rm sc} \equiv L/(1+ L/l_{\rm sc}^r) $ for the total scattering length.  

Below, we solve the Boltzmann equations 
(\ref{10}) and (\ref{20}) for large voltages 
$ eU \gg \hbar \omega $ and lengths $ L \gg l^r_{\rm sc} $ 
analytically by the help of two  
approximations. 
In the first  approximation we use in the electronic Boltzmann 
equation (\ref{10}) positional and momentum independent phonon distribution 
functions $ n(k,x) \approx \overline{n} $. We shall determine 
$ \overline{n} $ then similarly to (\ref{55}) where the energy average 
is taken over those energies where $ n(k,x) $ or $ f_L(k,x) $, 
$ f_R(k,x) $ are non-zero, respectively. Thus we neglect  
large (infinite) energy regions where the 
electron and phonon distribution functions are zero since 
they do not contribute to the current.  

In the calculation below, we find for the energy averaged 
phonon distribution functions  
\begin{equation} 
\overline{n}(x) = \frac{1}{2 (\epsilon_u-\epsilon_d)} 
\int^{\epsilon_u}_{\epsilon_d}  
 \! \!  d \epsilon \, n(2 k_L(\epsilon),x)+
n(2 k_R(\epsilon),x)  \label{62} 
\end{equation}     
where 
\begin{equation} 
\epsilon_u = \left\{ 
\begin{array}{ c c c}   
 eE x & \mbox{for} &  \pi \frac{e U l_{\rm sc}}{L}  \gg \hbar \omega \,, \\
  \frac{ \omega L}{\pi l_{\rm sc}} & \mbox{for} & 
\pi \frac{e U l_{\rm sc}}{L}  \ll \hbar \omega           
\end{array} \right.                          \label{63}       
\end{equation}
and 
\begin{equation} 
\epsilon_d = \left\{ 
\begin{array}{ c c c}   
 eE (x-L) & \mbox{for} &  \pi \frac{e U l_{\rm sc}}{L}  \gg \hbar \omega \,, \\[1mm]
  0 & \mbox{for} & 
\pi \frac{e U l_{\rm sc}}{L}  \ll \hbar \omega \,.          
\end{array} \right.                          \label{64}       
\end{equation}
$ \overline{n} $ is then determined by the average 
\begin{equation} 
\overline{n} = \frac{1}{L} \int^L_0 dx \; \overline{n}(x) \,.  \label{65}
\end{equation}

The second approximation is given by a linearization 
of the non-linear scattering terms in (\ref{30}) which can be identified 
by extracting the brackets in (\ref{30}). These terms are equal  to the terms
followed by setting in (\ref{30}) $ n^\nu $ equal to zero.    
To linearize these terms we should take care on the expansion points 
 $ \overline{f}_L $ where 
$ f_L = \overline{f}_L + \Delta f_L $ and similar for  $ f_R $. 
In a first crude approximation  we use in the following as the expansion point 
$ \overline{f}_L=\overline{f}_R=1 $ which are the 
boundary values (\ref{42}) for $ f_L $ and $ f_R $ on the non-zero 
momentum support of the electronic distribution function.    
Then we obtain 
\begin{align}
&\hspace{-1.6cm} f_R(k_R(\epsilon+\hbar\omega^\nu))[1-f_L(k_L(\epsilon))]
  \nonumber \\&
\hspace{-1.6cm} -
f_L(k_L(\epsilon))[1-f_R(k_R(\epsilon-\hbar \omega^\nu))]   \nonumber \\
& \approx f_R(k_R(\epsilon-\hbar \omega^\nu))-f_L(k_L(\epsilon))\,. \label{60}
\end{align}

\subsection{Current-voltage characteristic and electron distribution 
function} 
With the help of the approximations (\ref{65}) and (\ref{60}) one can 
solve (\ref{10}) with (\ref{30})  using Fourier methods. After a lengthy 
calculation carried out in Appendix A we obtain for the electron 
distribution function $ f_L(x,k_L(\epsilon))  $ (\ref{105}) with 
(\ref{110}) and (\ref{150}). The distribution function 
$ f_R(x,k_R(\epsilon))  $ is then given by $ f_L(x,k_L(\epsilon))  $  
with the help of the substitution (\ref{100}).  
The current voltage characteristic is given by (\ref{160})   
\begin{equation} 
I= 4 \frac{e}{h} \left[ \hbar 
 \omega B_1\left(\frac{1+{L}/{l^r_{sc}}}{\pi \overline{n}}
\right) +  \frac{e U}{\left(1+ {L}/{l^r_{sc}}\right)} 
\right]  
            \label{c160} 
\end{equation} 
where $ B_1(x) $ is defined in  (\ref{170}) and plotted in Fig.~6.   
By taking into account 
the regime $ \overline{n} \gtrsim 1 $, $ v_F \tau_{\rm ep} \approx 130 $ 
we obtain for $ L=300 $nm that $ B_1 \approx 0.5 $.     
By using $ l_{\rm sc} \approx 10.4$nm (this will be shown below 
by using Fig.~7)  
we obtain excellent agreement with  
the numerically determined current-voltage characteristic 
at high voltages shown in Fig.~3 and with experiments \cite{Javey1, Park1} 
measuring $ l_{\rm sc} \approx  10-11 $nm.   
Note that the first term in (\ref{c160}) corresponds to the y-axis value  
obtained by extending the high-voltage curves to this axis. 
For this we further note that $ 4 (e/h)  \hbar \omega  \approx 26.6 \mu $A.

Next, we discuss the current-voltage characteristic for large lengths, i.e. 
$ l_{\rm sc}/L \ll \tau_{\rm ep}/\tau_{\rm op} $ and low voltages  
$ \pi eU \ll \hbar \omega L/l_{\rm sc} $. Here, we obtain $ \overline{n} \approx 1.2 $ 
(see the discussion in Subsect. B.2) leading with Fig.~6 to 
$ B_1 \approx 1.3 $ for nanotube lengths $ L\approx 3000$nm. 
This leads to currents which are approximately two 
times the current values shown in Fig.~1.
Although this value is too large, the overall behavior of an 
approximatively length independent current for fixed bias voltage shown 
in the numerical calculation  
in Fig.~1, is also seen in the analytic calculation.     

Summarizing, from (\ref{c160}) we obtain a different behaviour 
of the current voltage characteristic in the regime 
$ \pi eU \gg \hbar \omega L/l_{\rm sc} $ where the second  term in
 (\ref{c160}) is the leading contribution to the current and the regime 
$ \pi eU \ll \hbar \omega L/l_{\rm sc} $ where the first  term is 
most relevant. The reason for this can be seen in (\ref{10}) 
and (\ref{30}). At low optical phonon scattering rates, i.e.  
$ \pi eU \gg  \hbar \omega L/l_{\rm sc} $, electron scattering 
takes primarily place from the upper part of the filled 
right moving band to the empty part of left moving band. For 
higher scattering rates $ \pi eU \ll \hbar  \omega L/l_{\rm sc} $ 
a large amount of electrons are also able to be scattered 
from the upper part of the filled right moving band 
to the filled part of the left moving band within  many scattering processes 
where now Pauli blocking prohibits this scattering. 
This Pauli blocking is only roughly  described by 
the linearized phonon scattering approximation (\ref{60}). This is 
the reason that in the large length regime we obtain less agreement 
between our numerical and analytical results in contrast to nanotubes 
of smaller lengths.

\begin{figure}
\begin{center}
\includegraphics[height=5cm,width=7.5cm]{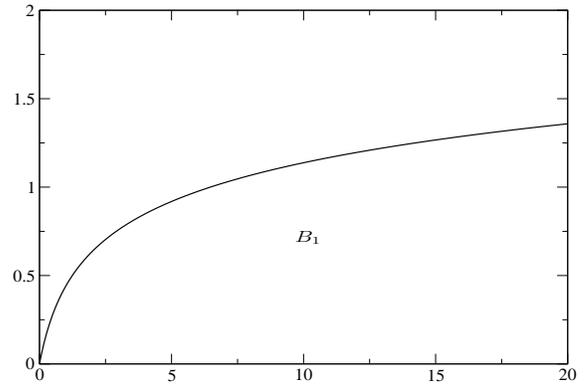}
\vspace*{0.5cm}
\caption{(Color online) 
We show the function $ B_1(x) $ defined in (\ref{170}).}
\end{center}
\end{figure}

\subsection{Phonon distribution functions} 

Next, we determine the phonon distribution function $ n(k,x) $ by solving 
(\ref{20}) with (\ref{40}) and (\ref{50}). We use here in our analytical 
calculation for simplicity 
the standard relaxation time approximation (\ref{50}) for the optical 
scattering term instead of the more complicated scattering term (\ref{1030})
which takes into account also the second generation phonons. 
As was discussed in Sect.~III the differences in the current-voltage 
characteristic are only minor when taking the effective optical 
phonon relaxation times $ \tau_{\rm op}=0.9 $ps for long nanotubes and 
$ \tau_{\rm op}\approx 9.1 $ps for short ones.  
We shall determine first the phonon distribution function $ n(k,x) $ 
 in the regime $\pi  e U \gg \hbar \omega L/l_{\rm sc} $.

\subsubsection{Phonon distribution function for   $ \pi eU \gg \hbar \omega 
L/l_{\rm sc} $}
  
We obtain in App.~A for 
$ n(k,x) $ (\ref{210}), (\ref{215})  and  (\ref{220}) for $ k>0 $ in 
the different 
parameter regimes (\ref{180}). 
For $ k<0 $ $ n(k,x) $ is given by $ n(-k,-x) $ (\ref{178}) where 
the $ K_1 $s are defined in (\ref{200}). From these equations 
we can calculate the energy averaged phonon distribution function 
$ \overline{n}(x) $. This function is given by (\ref{230}) with 
(\ref{240}), (\ref{242}) and (\ref{250}). 
From this function we obtain for the position 
averaged phonon distribution function $ \overline{n} $ (\ref{255})  with 
(\ref{260}), (\ref{262}) and (\ref{265}). This function determines effectively the 
averaged phonon distribution function by using the definition 
for $ \tilde{K}_1 $ (\ref{225}) in the regime $ {\cal R}_2 $ (\ref{180}). 
We solved this equation analytically below 
Eq.~(\ref{270}) in various nanotube length regimes. 
In Fig.~7 we show the numerical 
solutions for  $ \overline{n}$ and $ \tilde{K}_1 $ in the regime 
$ {\cal R}_2 $ as a 
function of the nanotube 
length $ L $. For a nanotube of length $ L=300 $nm we obtain 
$ l_{\rm sc} \approx 10.4 $ and $ \tilde{K}_1 \approx 0 $. For these values 
we obtain (\ref{290}) for $ \overline{n}(x) $ showing in fact a phonon 
distribution function which increases at the boundary of the nanotube.  
From (\ref{250}) we obtain that this behavior is even more pronounced 
for $  \tilde{K}_1 $-values larger than zero.

\begin{figure}
\begin{center}
\includegraphics[height=6cm,width=7.5cm]{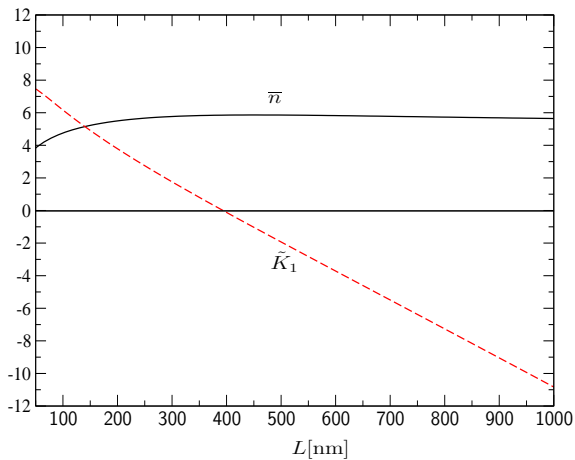}
\caption{(Color online) 
The (black) solid curve is given by the phonon distribution function 
$ \overline{n} $ as a function 
of the nanotube length $ L $
calculated with the help of (\ref{255}) with (\ref{260}), (\ref{262})  
and (\ref{265}).  
The (red) dashed curve shows $ \tilde{K}_1 $ (\ref{225}) as a function 
the nanotube length $ L $ in 
the parameter regime $ {\cal R}_2 $ (\ref{200}) 
using $ \overline{n} $.}
\end{center}
\end{figure}

\subsubsection{Phonon distribution function for   $ \pi eU \ll \hbar \omega L/l_{\rm sc} $}

In the following, we restrict ourselves to the regime 
where $ l_{\rm sc}/L \ll \tau_{\rm ep}/\tau_{\rm op} $ which is 
good fulfilled in the large length regime considered in Sect.~II. 
By taking into account (\ref{110}), (\ref{150}) and (\ref{100}) we obtain 
that $ n(k,x) \not= n^{\rm ac}_B $ only in the regime 
$ |\pi (\hbar v_F |k| /2 \omega) L/l_{\rm sc}| \lesssim 1 $ where 
we take into account that $ \overline{n} \lesssim 1 $ as will be shown  
immediately  below.

This leads to (\ref{310}) for $ n(k,x) $ and 
 (\ref{320}) for $ \overline{n}(x) $, $\overline{n} $ in this regime.  
By taking into account $ s^p 
\tau_{\rm op}/\tau_{\rm ep} \approx 4.76 $ we obtain 
$ \overline{n} \approx 1.2 $. Note this value is much larger than the 
numerical determined values shown in Fig.~2.

\section{The electron-phonon coupling induced frequency shift and 
lifetimes of optical phonons at high bias} 

By using the electron distribution functions $ f_L $ (\ref{105}) with 
(\ref{110}), (\ref{150}) 
and $ f_R $ (\ref{100}) we are now able to calculate the effect of 
a large bias voltage on the frequency shift and lifetimes of optical
phonons. We restrict our calculation to the Raman active zone 
center $ {\bf  \Gamma } $ phonons. In this section, we shall carry out a similar calculation as was done in  
Ref.~\onlinecite{Ishikawa1} for the  frequency shift and lifetimes
of optical phonons at zero bias voltage 
and temperature $ T=0 $. There,  
the electron distribution functions $ f_L $ and $ f_R $ are 
Fermi-functions. The retarded phonon 
self-energy at zero momentum is given by
\begin{equation} 
 \Pi^{L,T}(\omega)= \sum\limits_n 
\int\limits_{-1/2}^{1/2}dt   \,
[\Pi^{L,T}(n,\omega)-\Pi^{L,T}(n+t,0)] \label{500}
 \end{equation} 
where $ \Pi^{L,T}(n,\omega) $  is the self-energy 
contribution of the $ n$th electron band. Its value is given by  
\cite{Ishikawa1} 
 \begin{align} 
& \Pi^{L,T}(n,\omega)=   
-4 \sum\limits_{s,s'}  \sum\limits_{k}  dt 
\left(\frac{\beta \gamma}{b^2} \right)^2 \frac{ \hbar}{N M \omega^\Gamma} 
\label{510} \\
& \times \frac{1}{2}  \left( 1 \pm \frac{s s'[ \kappa^2[n] -k^2]}{
   \kappa^2[n] +k^2}       \right)  
\frac{f[\epsilon^s(n,k)]- f[\epsilon^{s'}(n,k)]}{
\hbar \omega-\epsilon^s(n,k)+\epsilon^{s'}(n,k)+ i0}  \nonumber 
\end{align}  
where the electron bands for metallic nanotubes are given by 
\begin{equation} 
\epsilon^s(n,k)=s \hbar v_F  \sqrt{
\kappa(n)^2 +k^2}
\label{525}  
\end{equation} 
with $ \kappa(n)=2 \pi n/L$ . $s=+1 $ for the energy levels in the 
conduction band $ \epsilon >0 $ 
and $ s=-1 $ for the energy levels in the valence band $ \epsilon <0 $.     
$ N $ is given by the number of unit cells and  $ M $ is the mass of a 
Carbon atom. The subtraction of the last term in (\ref{500}) is 
due to the fact that in order to calculate the frequency shift 
for nanotubes, we insert in the calculated expressions the known optical  
frequencies of graphene which then results in a double 
counting when we only use the first term in (\ref{500}) as self-energy 
\cite{Ishikawa1}. The valley degeneracy is here considered by a factor two in 
correspondence to similar expressions in Ref.~\onlinecite{Ishikawa1}. 
The upper sign 
corresponds to the self-energy of longitudinal phonons $\Pi^{L} $, 
the lower sign to the transversal ones $\Pi^{T} $  

In the case of the frequency shift of the longitudinal 
and transversal optical $ {\bf \Gamma} $ mode Ishikawa {\it et al.} use
the zero temperature Fermi-function for $ f[\epsilon^s(n,k)] $. 
The frequency shift $ \Delta \omega $ and broadening $ \Gamma $ is given by 
\cite{Ishikawa1} 
\begin{equation}
\Delta \omega = {\rm Re} [\Pi(\omega^\Gamma)] \quad , \quad   
\Gamma = - {\rm Im}[ \Pi(\omega^\Gamma)] \,.      \label{520}  
\end{equation} 
In Ref.~ \onlinecite{Ishikawa1} it is then shown that this leads 
to a good agreement of the theoretically calculated frequency shifts and 
broadenings by using (\ref{500})-(\ref{520}) and the experimentally 
determined ones using Raman spectroscopic methods.
In the following, we carry out a similar calculation for the case of 
the electron system under high bias voltage. For this we use for 
$ f[\epsilon^{s}(n,k)] $ in (\ref{510}) for the lowest energy band, 
i.e. $ n=0 $, the distribution functions $ f_L $ and 
$ f_R $ calculated in the last section. For the higher bands $ n \not=0 $ 
we shall use the Fermi-function since we did not take into account higher band 
excitations in Sect.~IV being negligible in the considered voltage regime.  
\begin{equation} 
f[\epsilon^s(n,k)]= \left\{ \begin{array}{c c c} 
                              f_L(k)& {\rm for} & n=0 \quad  \& 
\quad  
k\cdot s  <0  \\
                              f_R(k) & {\rm for} & 
n=0 \quad  \& 
\quad  
k \cdot s >0  \\ 
                              n_F(\epsilon^s(n,k)) & {\rm for} & n\not=0 
                           \end{array}  \label{530} \right. .
\end{equation} 

Expression (\ref{510}) for the phonon self-energy corresponds to
a current-current Green's function loop being the lowest 
order approximation for the phonon self-energy. When using dressed Green's 
functions by taking into account the electron-phonon 
interaction and also the external electric field 
one has to use non-equilibrium 
Green's function techniques in order to get the corresponding  
loop expression \cite{Haug1} for the retarded phonon self-energy. 
The relevant Green's functions in the loop are given by lesser $ G^< $ 
and greater 
Green's functions $ G^> $. We now express these 
Green's functions by the spectral function and use the 
quasi-particle approximation \cite{Haug1} 
established for deriving the Boltzmann equation. This leads to 
(\ref{500}), (\ref{510}) with (\ref{530}). 

It is well known that, by using the free Green's 
function in loops as was done by Ishikawa {\it et al.} \cite{Ishikawa1},  
this approximation is conserved \cite{Baym1} which means that 
the charge-current response functions corresponding to the loop fulfill 
the continuity equation \cite{Baym1}. By using a more general 
dressed Green's function in the loop one needs also vertex 
corrections in order to fulfill the continuity equation. 
It is straight forward to show that the current-density 
correlation functions corresponding to (\ref{500}), (\ref{510})  
with (\ref{530}) which 
consist of dressed Green's function in loops with the 
additional quasi-particle approximation does indeed fulfill 
the continuity equation. This justifies approximation (\ref{500}), (\ref{510}) 
with (\ref{530}) 
for the phonon self-energy at zero momentum to calculate the phonon 
self-energy under high bias voltage.              

In the following, we use the abbreviations: 
\begin{eqnarray}
\tilde{\omega} & = & \omega \frac{\pi D}{2 \pi v_F} \,, \label{540} \\
\alpha(D)  & = & \frac{27}{\pi} \beta^2 \left(\frac{2 (\hbar v_F)}{\sqrt{3}} 
\right)  
\frac{\hbar^2}{2 M a^3} \left(\frac{1}{\hbar \omega^\Gamma}\right)^2 \frac{a}{\pi D }
\,. 
  \label{545} 
\end{eqnarray}
Here $ a $ is $ \sqrt{3} $ times the equilibrium bond length.   
With the knowledge of the 
parameter $ \tau^\Gamma_{\rm ep} = 538 $fs determined by density functional 
methods for a nanotube of  diameter $ 2$nm 
\cite{Lazzeri2}
we are able to determine $ \beta $ and thus $ \alpha(D) $.        
By  using that $ \tau^\Gamma_{\rm ep} $ is more generally 
proportional to the diameter 
of the nanotube \cite{Lazzeri1} we obtain from (\ref{545}) that 
\begin{equation} 
\alpha(D) \approx 0.1 \, a/\pi D  \,.  \label{548} 
\end{equation}    
  
With the help of (\ref{500}), (\ref{510}) and (\ref{530})   
by using the abbreviation 
$ \Pi^L(\omega) = \Pi^L_{U=0}(\omega)+\Pi^L_{U>0}(\omega) $  
we obtain 
\begin{eqnarray} 
\! \! \! \Pi^T(\omega) \! & = & \! \alpha(D) \omega^\Gamma \! \! \left(2 - \frac{\pi^2}{9} 
\tilde{\omega}^2 \right) \,, \label{550} \\
\! \! \!  \Pi^L_{U=0}(\omega)  \! & = & \! \alpha(D) \omega^\Gamma \! \! \left( \! \! 2 \ln\frac{ 2\pi \tilde{\omega}}{ e} 
- \frac{\pi^2}{18} \tilde{\omega}^2 \! - \! i \pi \! \! \right) \,. \label{560}
\end{eqnarray} 
$ \Pi^L_{U>0} $ is a correction factor to the zero bias 
self-energy $\Pi^L_{U=0}$ (\ref{560}). By taking into account that  
$ \Pi^T(n=0,\omega) =0 $ (\ref{510}) we obtain that this factor is 
of similar order as the zero-bias 
self-energy only in the 
longitudinal sector for $ e U \ll \hbar v_F 2/D $.
Note that we restrict here ourselves 
also to the 
regime $ e U \ll \hbar v_F 2/D $ as was done in  the last  
sections by taking into account that excitations to higher bands 
are negligible.   
We obtain for the self-energy $ \Pi^L_{U>0} $ 
\begin{equation} 
\Pi^L_{U>0}(\omega, x)= \Pi^L_1(\omega, x)+\Pi^L_2(\omega, x)  \label{565} 
\end{equation}   
with 
\begin{align} 
& \Pi^L_{1}(\omega, x) = \alpha(D) \omega^\Gamma \left\{ 
2 \, C(\tilde{x})+ 
2  \ln\left(\frac{2 e U}{\hbar \omega} \right)
+ i\pi \right\}
 \,, \label{570}  \\ 
& \Pi^L_{2}(\omega, x) \approx - \alpha(D) \omega^\Gamma \frac{1}{2}  
 \left\{
(1-\tilde{x}) 
\ln\left(\frac{|\tilde{x} 
 + \frac{\hbar \omega}{e U \pi}
\frac{L}{
 l_{\rm sc}
}|}{\tilde{x} }   \right)   \right.   \nonumber \\
& \left.   +  \tilde{x} 
 \ln\left(\frac{|1-\tilde{x} -\frac{\hbar \omega}{e U \pi} \frac{L}{
l_{\rm sc}}|} {1-\tilde{x}} \right)        \right\}  + 
\left\{ \frac{\hbar \omega}{e U \pi}\frac{L}{
l_{\rm sc}} \rightarrow  - \frac{\hbar \omega}{e U}  \overline{n} \right\} ,
      \label{580}      
\end{align}
where 
\begin{equation} 
C(x) \equiv x \ln(1-x)+ (1-x) \ln(x) \,.   \label{585} 
\end{equation}
and $ \tilde{x}=x/L $.    
The last line in (\ref{580}) means that 
we have to add the foregoing 
expressions with the substitution 
$ (\omega/\pi) L/l_{\rm sc} \rightarrow  - \omega \overline{n} $.
Here $ \Pi^L_{1} $ is the self-energy part calculated by the help 
of $ f^{t=0} $ (\ref{110}) in (\ref{510}). $ \Pi^L_{2} $ is the result for 
$ f^{t\not=0 } $ (\ref{150}). 
In (\ref{580}) 
 we carry out for the logarithmic term in the electron distribution 
function (\ref{130}) the approximation that 
we set this term constant over the range $ \pi |\epsilon- eEx_1| l_{\rm sc} 
/\hbar \omega L < 1 $ with its value at $ |\epsilon- eEx_1|=0 $ and zero 
elsewhere. This approximation leads to the logarithmic singularities in 
(\ref{580}). They 
are  softened when using the exact functions 
$ f_L^{t \not=0} $ (\ref{150})  without 
approximation but this treatment has the disadvantage 
that we would  not obtain analytical 
results. We should additionally mention that we neglect 
those imaginary terms in (\ref{570}) and (\ref{580}) which exist only  
in small regions in position space 
of length $ \Delta x= L \hbar \omega/e U $. These correspond  
to regions where the ln-terms
in (\ref{570}) and (\ref{580}) gets singular.        

By taking into account the results in (\ref{560}), (\ref{565}), 
(\ref{570}) and (\ref{580}) 
we obtain the surprising result that the imaginary part of the longitudinal 
phonon self-energy and thus the level broadening  
vanishes 
\begin{equation} 
 {\rm Im} [\Pi^L(\omega,x)]=0      \label{587} 
\end{equation} 
in the high voltage-bias regime $ e U \gg \hbar \omega $ in contrast to 
the case of no bias (\ref{560}). The reason for this vanishing is seen 
in (\ref{510}). We only obtain an imaginary part for the 
phonon self-energy when there is a substantial changing of the 
electron distribution function in the energy range $ |2 \epsilon| < 
\hbar \omega $. 
In the high bias regime the electron distribution 
function (\ref{530}) becomes constant in this range which is not the 
case for the Fermi-function in the zero-bias system.    
 
\subsection{Position averaged self-energy} 

Finally we calculate the position averaged 
self-energy 
$ \overline{\Pi}_i^L= \int dx  \Pi_i^L(\omega,x)/L $. 
We obtain 
 \begin{align} 
& \overline{\Pi}^L_1(\omega)  = 
 \alpha(D) \omega^\Gamma\left[-3 + 
2 \ln \left( \frac{2 e U}{\hbar \omega}
\right) + i \pi \right] \label{590}  \\
& \overline{\Pi}^L_2(\omega)  \approx  \alpha(D) \omega^\Gamma 
\left[F_1\left(\frac{\hbar \omega}{\pi e U} \frac{L}{l_{\rm sc}}\right) 
+F_1\left(\frac{\hbar \omega}{e U}\overline{n}\right) \right]
           \label{600}
\end{align}  
with 
\begin{eqnarray} 
F_1(x) & = &  \frac{1}{4} (x+1)^2 \, [\ln(|x|)-\ln(|1+x|)]-\frac{1}{4} 
\ln(|x|) 
\nonumber \\
& & + (x \rightarrow -x) 
   \label{605} 
\end{eqnarray}   
 \comment2{is this the correct bracket setting?}
We show in Fig.~8 the function $ F_1(x) $. 
In the regime of very high voltages 
$ e U l_{\rm sc} / L \gg  \hbar \omega/\pi $ we obtain that 
$\overline{\Pi}^L_2(\omega) $ is negligible in comparison to 
$\overline{\Pi}^L_1(\omega) $. For nanotubes of diameter $2 $nm we have 
$ \tilde{\omega}_0 \approx 0.35$. Thus we obtain that 
$ {\rm Re} [\Pi^L_{U=0}(\omega^\Gamma)] \approx 
-0.46 \, \alpha(D) \omega^\Gamma $ leading to the result that 
$ {\rm Re} [\Pi^L(\omega^\Gamma)] \approx  
\alpha(D) \omega^\Gamma[-3.48+2 \ln(2 e U/\hbar \omega^\Gamma)] $. 
This means that we obtain in the high 
voltage regime for $ e U/\hbar \omega^\Gamma \gtrsim 1.8 $  
a positive frequency shift for the longitudinal 
optical phonon frequency at the $ {\bf \Gamma} $ point. 
On the other hand, by taking into account that $ \lim_{x \to \infty} 
F_1(x)= -1/2 \ln(x) $ we obtain that for fixed $ e U/ \hbar \omega^\Gamma 
$ and 
in the large length limit $ L/l_{\rm sc}  \to \infty $ 
where $ \overline{n} \lesssim  1 $, that the frequency shift is negative. \\

\subsection{Line-shape of Raman signal} 

From the considerations above it is not clear how the actual line-shape 
of the Stokes or anti-Stokes signal looks like in an actual Raman scattering 
experiment. From above we obtain first 
that the Raman signal corresponding to the response on   
transversal phonon mode excitations denoted by 
$ G^+ $ is not changed from the zero bias result.   
This is not true for the $ G^- $ Raman mode corresponding to 
the response on longitudinal optical phonons. 
In order to get a better insight into the actual Raman line-shape  we 
assume that the incident laser light illuminates the nanotube continuously 
over the whole width. In typical Raman experiments the scattering of 
the electron  system with the incident light is dominated 
by the resonant scattering of a valence band 
and conduction band of fixed index $ n>0 $ (\ref{525}) 
\cite{Dresselhaus1}. A further enhancement of the signal 
is reached when electrons from the band edges are scattered. 
This is 
due to the 
van Hove singularities of the density of states at this region leading 
not until then to the opportunity of measuring phonon and electronic 
properties of single nanotubes. 
Further let us assume in a first approximation 
that the phonon 
distribution function $ n^\Gamma(x) $ (\ref{55}) is homogeneous 
over the nanotube width which was 
also assumed in our analytical calculations in Sect. III. 
This leads us to the 
conclusion by taking into account that only the electron distribution 
function of the lowest electronic energy band $ n=0 $ is changed due to 
the large bias voltage that we can determine at least approximately 
the line-shape of an actual 
 Raman signal by the position average  over the individual 
Raman signals.

In the following, we use the abbreviation 
\begin{equation} 
F_2(x)=\int^1_0 dy \;   \delta(C(y)-x)       \label{610} 
\end{equation}
in order to calculate the line-shape of the Raman signal. 
We restrict ourselves 
 in the following to the most important high bias regimes 
$\pi e U l_{\rm sc}/L \gg \hbar \omega^\Gamma $ and to the large length regime 
where $\pi e U l_{\rm sc}/L  \ll \hbar \omega^\Gamma  $, but  
$eU \gg \hbar \omega \overline{n}$ 
where we obtain simple expression for the Raman signal.    
By taking into account that $ \rm Im[\Pi^L] =0 $ 
we obtain for the line-shape of Stokes and anti-Stokes signal $ I_{\rm ep}  $ 
by using (\ref{570}), (\ref {580}) and (\ref{610})
\begin{equation} 
I_{\rm ep} (\omega) \! \! \propto\! \!  \left\{  
\begin{array}{c }  
\! \! \!  \!\frac{1}{2} F_2 \! \left[\frac{\omega 
- \omega^\Gamma- {\rm Re}[\Pi^L_{U=0}(\omega^\Gamma)]}{ 2 \alpha(D) \omega^\Gamma} 
\! - \! \ln\!\left(\frac{2 e U}{\hbar \omega^\Gamma}\right)\right] \!  \mbox{\rm for}    
 \frac{e U}{\hbar \omega^\Gamma}\!  \gg \!\!  \frac{L}{\pi l_{\rm sc}}   \\  
\hspace*{-1.8cm}  \frac{2}{5} F_2\left[\frac{\omega - \omega^\Gamma- {\rm Re} 
      [ \Pi^L_{U=0}(\omega^\Gamma)]}{ (5/2) \alpha(D) \omega^\Gamma} 
-\frac{4}{5} \ln\left(\frac{2 e U}{\hbar \omega^\Gamma}\right) \right. \\ 
\left. + \frac{1}{5} \ln\left(\frac{ \hbar  \omega^\Gamma}{\pi e U} 
\frac{L}{l_{\rm sc}} \right)   
\right] \quad   \mbox{\rm for} \quad   
   \overline{n} \ll \frac{e U}{\hbar \omega^\Gamma} \ll  \frac{L}{\pi l_{\rm sc}} .   
\end{array}  \right.           \label{620} 
\end{equation}      
We show on the right hand side in  Fig.~8  the function $ F_2$. 
Due to the the maximum of 
$ C(x) $ at $ x=1/2 $ with $ C(1/2)= -\ln(2) $ we obtain that 
$ F_2(x) $ 
is singular at $ x= - \ln 2 $.  This singularity 
leads to a sharp edge of the Raman spectrum at the corresponding 
frequency according to (\ref{620}). 

By taking into account also the broadening of the phonon modes due to 
the phonon-phonon scattering we obtain that 
the signal of the actual Raman mode is a convolution of 
$ I_{\rm ep} (\omega) $ (\ref{620}) 
with a Lorentzian which has a negative 
frequency shift and broadening corresponding to the 
phonon-phonon interaction contribution to the phonon self-energy. 

\subsection{Discussion} 

Next we compare our findings with recent experiments. 
In Refs.~\onlinecite{Bushmaker1, Deshpande1} the optical phonon 
lifetimes and frequency shifts are measured for {\it suspended} 
metallic carbon nanotubes at high bias voltage using Raman spectroscopy.
These investigations did not find a response of the $ G_- $ and 
$ G_+ $ modes corresponding to the longitudinal  and transversal
optical $ {\bf \Gamma} $ modes for all measured nanotubes 
where only the former mode couples to the electron system at high bias.  
In fact, it is not well understood which mode response to the bias 
for a certain nanotube.
The nanotube phonon temperature was then  measured by two different methods 
either by the Stokes/anti-Stokes intensity ratio being a function of 
the phonon Boltzmann factor of the corresponding mode or by 
mode softening where one compares the softened mode with the 
known temperature softening which is caused 
by phonon-phonon  scattering. Astonishing, 
that both methods find phonon temperatures 
which are in good accordance. 
They did not provide  any hint for a different behavior 
which we found  theoretically above. 
 
Such a different behavior was in fact seen by Oron-Carl {\it et al.}
\cite{Oron-Carl1} for metallic nanotubes on a SiO$_2 $ {\it substrate}.
They found a mode hardening and a reduction of the linewidth at high
bias voltage in comparison to the zero bias voltage values.  A real
quantitative comparison of our results with the experimental finding
is difficult here since the overall broadening and hardening is 
the additive effect of phonon-phonon scattering leading to
softening and broadening of the mode and the hardening effect and a
linewidth reduction due to the electron-phonon coupling.  That this
non-thermal effect was not seen in free-standing nanotubes could be
explained by the fact that the heating of many of the crystal modes
are much easier for phonons in a suspended crystal in comparison to
phonons on a substrate since these phonons could travel especially in
long nanotubes substantially until they diffuse out of the system. This
is the reason for the experimental finding 
that large nanotubes show a bias response for
the $ G_+ $ mode and the $ G_- $ mode in general \cite{Deshpande1}.
During this travelling the electron-scattered optical phonons  
are then thermalized by scattering with acoustic phonons.  
A true understanding of the different behavior of the frequency 
shift and broadening between suspended nanotubes 
and nanotubes on a substrate requires 
a new theory for suspended nanotubes being
out of the scope of this work.  
   
\begin{figure}
\begin{center}
\includegraphics[height=6cm,width=7.5cm]{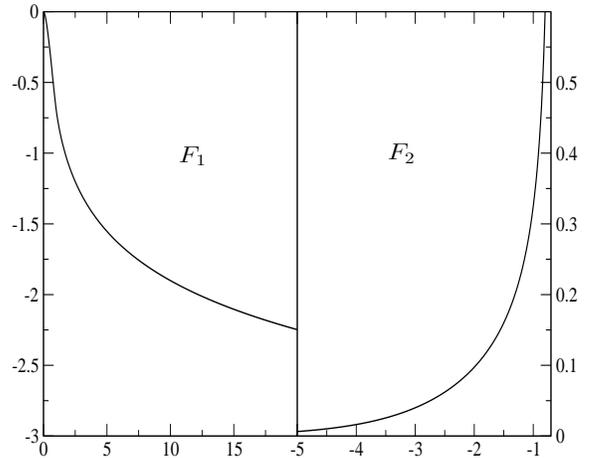} 
\vspace*{0.7cm} 
\caption{(Color online) 
The left panel shows the function $ F_1(x) $ 
(\ref{605}). On the right hand side we show the 
Raman lineshape function $ F_2(x) $ defined in (\ref{610}).}
\end{center}
\end{figure}
      
\section{Summary}
In this paper we have used the coupled Boltzmann equations for electrons and 
optical phonons to calculate the current-voltage characteristic of carbon 
nanotubes  lying on a substrate under high bias voltage. First we have studied
 the coupled electron-phonon Boltzmann system. By taking into account 
the  electron-phonon relaxation time of Ref.~\onlinecite{Lazzeri1}
which agrees well with experiments, we have determined by numerical fitting 
the relaxation times of the optical phonons
in the single-mode relaxation time approximation. 
These are much longer for short nanotubes below 
$ 1 \mu$m than for large ones. 
The result was obtained 
 by fitting our numerically determined 
current-voltage curves  with the experiments. 
For short nanotubes, this time did not agree
with 
experimental findings. 
In Sect III we went beyond the single time 
relaxation approximation by taking into account also lower lying 
secondary phonons in the Boltzmann equation which leads us to 
the conclusion that the phonon relaxation shows a bottleneck in the 
sector of acoustic phonons for short nanotubes,
but not
for long nanotubes. We have explained this by the fact that due to the 
phonon velocity of the lower-lying acoustic phonons, these phonons  
are redistributed in such a way that at least locally 
a bottleneck is created for short 
nanotubes. This leads  to a plug in the relaxation path of the optical 
phonons.   

In Section IV, we have 
considered an analytical solution of the Boltzmann system 
where we first linearize the electronic equations  
and use further 
the assumption of a constant phonon distribution function 
to solve them in the electronic sector. We have 
compared our results for the current-voltage characteristic 
and the phonon distribution function with the numerical findings 
of  Sect.~II, 
Sect.~III and the experimental results. 
We  find an 
especially good agreement in the high-voltage, small length regime. 
Our analytical theory 
provides
 us with the electron and phonon distribution 
functions as a function of position and momentum. This opens 
the possibility to calculate the optical phonon broadening and frequency shift 
due to the coupling of the phonon system to the bias driven 
electronic system. For zero bias this coupling 
is the dominant contribution for both quantities

We have then calculated in Sect. V in a charge-current conserved way 
the broadening and frequency shift of the zone-center optical phonons. 
We find that the
phonon-level broadening, determined at zero bias voltage 
mainly by the  electron-phonon coupling, vanishes at large voltage.  
The vanishing was explained
by a smoothing of the electron 
distribution function 
over the Fermi-level at non-zero bias voltage.   
For very large voltages and small nanotubes, we found a 
positive frequency  shift. 
Note that this behavior of a vanishing broadening and positive 
frequency shift is in contrast to the self-energy 
contribution of the phonon-phonon interaction 
to the lower lying hot acoustic phonons. 
This contribution to the self-energy led to a negative frequency shift
and an additional broadening at higher bias due to a temperature 
increase of the lower lying acoustic phonons. 
 
It was in fact just recently 
found experimentally 
by Raman-scattering, that the level broadening 
decreases  and  the frequency shift increases for the zone-center optical 
phonons \cite{Oron-Carl1} for increasing bias voltage.   
In contrast to the very high voltage regime, we found a negative frequency 
shift at moderate bias voltage and large nanotubes.

\acknowledgments
The authors acknowledge the useful discussions with 
V.~Bezerra, M.~Lazzeri, C.~Auer, 
F.~Sch\"urrer and C.~Ertler.  
We further acknowledge the support provided by Deutsche Forschungsgemeinschaft
under grant KL 256/42-3.

\begin{appendix} 

\section{Analytic calculation of the current-voltage characteristic} 

In the following, we carry out the calculation of the 
current-voltage characteristic and the electron-phonon 
distribution functions for $ e U \gg \hbar \omega $ 
explicitly by using (\ref{10})-(\ref{50}). To solve the system of 
equations we use the approximations (\ref{65}) and (\ref{60}). 
  
\subsection{Current-voltage characteristic} 

In the following calculation we go in Fourier-space 
$f_{L/R}(k_{L/R},x)=1/(2 \pi)^2 \int dk  dt \hat{\hat{f}}_{L/R}(t,k) 
\exp[i (-k x \mp t  v_F \hbar k_{L/R})] $ and use further the function 
$ \hat{f}_{L/R} $ 
 defined by $f_{L/R}(k_{L/R},x)=1/(2 \pi) \int dt  \hat{f}_{L/R}(t,x) 
  \exp[\mp i (t v_F \hbar k_{L/R})] $. 
With the help of      
\begin{equation}
st= \sqrt{1-\left[\frac{2 \overline{n} \cos(\omega t) + \exp(-i \omega t)}{2 
\overline{n}+1}\right]^2}        \label{70} 
\end{equation}   
we obtain from (\ref{10}) 
\comment2{Hier und danach Klammerhierarchie neu machen}
\begin{align}
& \hat{f}_L \!= \! \frac{\hat{n}_F(t)}{De} \Bigg\{ 
e^{-i e E (x-L)\, t/\hbar  }  
\! \left[ 
 st \cosh\left(\frac{|x|}{l_{\rm sc}^r}  \, st\right) +      \right. \nonumber \\
&  \left. 
\sinh\left(\! \frac{|x|}{l_{\rm sc}^r}  \, st \! \right) \right] \! \! \! + \! 
e^{-i e Ex \, t/\hbar } 
 (1-st^2)^{1/2} \!  \sinh\left(\! \! \frac{|x-L|}{l_{\rm sc}^r} st\right) 
\! \! \Bigg\},  
\label{80}
\end{align} 
with 
\begin{equation} 
De= st \, \cosh\left(\frac{ L}{l_{\rm sc}^r}\,  st\right)+ 
   \sinh\left(\frac{L}{l_{\rm sc}^r}  \, st\right)    ,     \label{90} 
\end{equation} 
where $ \hat{n}_F(t) $ is the Fourier-transform of the Fermi-function, 
i.e.   
$ \hat{n}_F(t)= \int d \epsilon  n_F(\epsilon) \exp[-i t \epsilon] =
+i  /(t+ i \delta)  $.
The function $ f_R $ is given by (\ref{80}) and (\ref{90}) with the 
substitution 
\begin{equation} 
f_R=f_L[x \rightarrow x-L , x-L  \rightarrow x]\,.         \label{100}     
\end{equation} 
We note that one has to take into account  during this substitution 
the absolute value signs 
in the cosine hyperbolic and sinus hyperbolic arguments in (\ref{80}).

In the following, we evaluate (\ref{80}) by the help of 
a residuum integration.   
With 
\begin{equation} 
f_L=f^{t=0}_L+f^{t\not=0}_L       \label{105} 
\end{equation}
we obtain for the electron 
distribution function $ f^{t=0}_L
$ due to the residuum at $ t=0 $  
\begin{align} 
& f^{t=0}_L = 
[1- n_F(\epsilon-e E (x-L))]  \nonumber \\
 &      +[n_F(\epsilon- e E x)-n_F(\epsilon-e E (x-L))]
\frac{|x-L|}{l^r_{\rm sc}} \frac{1}{1+ \frac{L}{l^r_{\rm sc}}}\,.   \label{110}
\end{align} 
$ f^{t=0}_R  $ is given by  (\ref{110}) with the substitution (\ref{100}).
The  $ t \not=0 $ singularities are given by the zeros of  $ De $. 
For $ L/l^r_{sc} \gg  1 $ we  obtain for these zeroes
\begin{eqnarray} 
  st & \approx & i \frac{\pi n}{\left(1+\frac{L}{l^r_{\rm sc}}\right)}      
\qquad  \mbox{for}  \qquad |st| \lesssim 1   \,,  
 \label{122}  \\
  st & \approx & i \left(\pi n -\frac{\pi}{2}\right) \frac{l^r_{\rm sc}}{L}
 \qquad \mbox{for}  \qquad |st| \gtrsim 1    \,.  
\label{126}            
  \end{eqnarray}

With the help of 
\begin{align}
& A_1(x_1,x_2) =  \frac{1}{2 \pi i}
 \Bigg\{  \left[1-n_F\left(\epsilon - e E x_1\right)\right]       
\nonumber   \\
& - n_F\left(\epsilon - e E x_1  \right) 
\exp\left(-\frac{1}{\overline{n}} \frac{1}{\omega} 
 \left|\epsilon - e E x_1 \right|\right) \Bigg\}   
\nonumber \\
&  \times \! \! \sum_{\pm} \! \! \pm \ln \! \! \left\{ \! \! 1+\exp\left[- 
\frac{\pi}{\left(1+\frac{L}{l^r_{\rm sc}}\right)} 
\left(\frac{  \left|\epsilon-e E x_1 \right| }{\omega} 
 \pm  i \frac{|x_2|}{l^r_{\rm sc}} \right)\right]\right\},   \label{130} \\
& A_2(x_1,x_2) = \frac{1}{2 \left(1+\frac{L}{l^r_{\rm sc}}\right)} 
\Bigg\{  \left(1-n_F\left(\epsilon - e E x_1\right)\right)       
\nonumber   \\
& - n_F\left(\epsilon - e E x_1 \right) 
\exp\left(-\frac{1}{\overline{n}} \frac{1}{\omega} 
 \left|\epsilon - e E x_1\right|\right) \Bigg\}   
\nonumber \\ 
& \times  \! \!   \sum_{\pm} \! \!  
\left\{\! 1+  \exp\left[ \frac{\pi}{\left(1+\frac{L}{l^r_{\rm sc}}\right)} 
\left(\frac{\left|\epsilon - e E x_1  \right|}{\omega} 
\pm   i \frac{|x_2|}{l_{\rm sc}^r}\right)\right] \right\}^{-1}          \label{140}
\end{align} 
we obtain for $ f^{t \not= 0} $ due to the singularities in the regime 
$ |st| \lesssim 1 $ (\ref{122})  
\begin{equation} 
f^{t \not=0}_L=A_1(x,x-L)+A_1(x-L,x)+A_2(x-L,x) .      \label{150} 
\end{equation}    
$ f^{t \not=0}_R $ is given by $ f^{t \not=0}_L $ with the substitution 
(\ref{100}). 

Finally, we have to discuss the contribution of singularities 
$ |st| \gtrsim 1 $ (\ref{126}) to $ f^{t \not= 0}  $.   
Due to the Fourier-exponent, these terms are exponentially suppressed  
beyond a small energy strip of range 
$ | \Delta \epsilon|/\omega \gtrsim 1 $ in comparison to terms calculated with 
residua $ |st| \lesssim 1  $. From the discussion above and the following 
discussions we obtain that 
these terms are not relevant in the range  
$ e U/\hbar \omega  \gg 1$, $ L \gg l^r_{\rm sc} $ for the determination of the 
phonon distribution function. We also note without an explicit 
calculation here but which can be shown with similar methods as for the 
current calculation of the $ |st| \lesssim 1 $ singularities below that 
the current from the $ |st| \gtrsim 1 $ singularities is negligible 
in comparison to the current calculated 
from the singularities $ |st| \lesssim 1 $. This current contribution 
will be calculated explicitly in the following.   
 
The current $ I $ can be calculated by the help of   
$ I = (4 e/h) \int d\epsilon (f_R- f_L) $. This is best done by  
carrying out the integration at the boundaries $ x=0 $ 
or $ x=L $ of the nanotube.  
From (\ref{110}) and (\ref{150}) and the corresponding expressions for 
$ f_R $, we obtain the following 
current-voltage characteristic
\begin{equation} 
I= 4 \frac{e}{h} \left[ \hbar 
 \omega B_1\left(\frac{1+\frac{L}{l^r_{sc}}}{\pi \overline{n}}
\right) + e \frac{U}{\left(1+ \frac{L}{l^r_{sc}}\right)} 
\right]  
            \label{160} 
\end{equation} 
with 
\begin{equation} 
B_1(x)=\frac{1}{2 \pi} \left[H\left(\frac{x}{2}-1\right)- 
     H\left(\frac{x}{2}\right)+\ln(4) +2
         H(x)\right]      \label{170}
\end{equation}    
where $ H(x)=\gamma+ \Psi(x+1) $ and $ \gamma $ is the 
Euler-Mascheroni constant,$ \Psi(x) $  is the digamma function.
We show in Fig.~6 the function $ B_1(x) $.

\subsection{Phonon distribution function}

Next we calculate the phonon distribution function 
$ \overline{n}(x) $ (\ref{62}). 
We carry out this calculation only in the leading order in 
$ l^r_{\rm sc}/ L $. In order to simplify our notation we define 
$ \tilde{x} \equiv x/L $.
Next we calculate from  (\ref{110}), (\ref{150}) and the corresponding 
equation for $ f_R $ (\ref{100}) the phonon distribution function 
in the regime $ \pi eU \gg \hbar \omega L/l_{\rm sc}  $ 

\subsubsection{Phonon distribution function for   $ \pi eU \gg \hbar \omega L/l_{\rm sc} $}

In the following, we only have to determine the $ k>0 $ 
phonons since we find from  symmetry arguments that  
\begin{equation} 
 n(k,x)= n(-k,-x)  \,.                    \label{178} 
\end{equation} 
To simplify our notation we define the following regimes for $ k >0 $ 
\begin{eqnarray} 
{\cal R}_1 & : &  \tilde{x}< \frac{1}{2} \, ,  e U (1-\tilde{x})  >  
\frac{v_F \hbar k}{2} >e U \tilde{x} \,, \label{180} \\
{\cal R}_2 & : &  \tilde{x}>\frac{1}{2}\, ,\,  e U \tilde{x}  >   
\frac{v_F \hbar k}{2}>e U 
(1-\tilde{x}) \,,  \nonumber \\
{\cal R}_3 & : &  \tilde{x}<\frac{1}{2} ,  e   U \tilde{x}  >   
\frac{v_F \hbar k}{2}
\; \mbox{or} \; \tilde{x}>\frac{1}{2} ,    e U (1-\tilde{x})  >   
\frac{v_F \hbar k}{2} .  \nonumber 
\end{eqnarray}

We obtain for the phonon-Boltzmann equation (\ref{40}) by 
using (\ref{110}), (\ref{150}) and (\ref{100})
\begin{equation} 
\frac{\partial}{\partial_x} n(k,x) = K_1  n(k,x) + K_2  \label{190} 
\end{equation} 
 with
\begin{eqnarray}
{\cal R}_1 & : & K_1= - \frac{1}{v_{\rm ph}}\left(\frac{s^p}{\tau_{\rm ep}} 
\frac{l_{\rm sc}}{L}+ \frac{1}{\tau_{\rm op}}\right)  
\,, \, K_2 = \frac{s^p \tilde{x} }{v_{\rm ph}\tau_{\rm ep}} ,\nonumber \\
{\cal R}_2 & : & K_1= - \frac{1}{v_{\rm ph}}\left(- \frac{s^p}{\tau_{\rm ep}} 
\frac{l_{\rm sc}}{L} +\frac{1}{\tau_{\rm op}}\right)
\,, \, K_2 = \frac{s^p (1-\tilde{x})}{v_{\rm ph}\tau_{\rm ep}},  \nonumber \\
{\cal R}_3 & : & K_1= - \frac{1}{v_{\rm ph}\tau_{\rm op}}
\,, \, K_2 = 2 \, \frac{s^p \tilde{x}(1-\tilde{x})}{v_{\rm ph}\tau_{\rm ep}}. 
\label{200} 
\end{eqnarray}    
By taking into account the boundary conditions $ n(k,0)=0 $ and 
further $ v_{\rm ph} \tau_{\rm op }\ll L $ we obtain for 
$ n(k,x) $ in $ {\cal R}_1 $ and $ {\cal R}_3 $ 
\begin{eqnarray} 
{\cal R}_1 & : & n_{{\cal R}_1} (k,x)= -\frac{s^p L }{v_{\rm ph}\tau_{\rm ep}}  \frac{1}{K_1 L} 
         \tilde{x}  , \label{210} \\  
{\cal R}_3 & : & n_{{\cal R}_3}(k,x)=-
\frac{s^p L}{v_{\rm ph}\tau_{\rm ep}} \frac{1}{K_1 L} 2 \, \tilde{x}(1-\tilde{x}).  \label{215}
\end{eqnarray}
We point out 
here that according to (\ref{200}) 
the expressions  for $ K_1 $ in (\ref{210}) and (\ref{215}) 
are different in the different regimes (\ref{180}).   
We obtain in these regions the same solutions $ n(k,x) $  of (\ref{20}) 
as we would set immediately $ v_{\rm ph}=0 $ from the beginning.
This is not true for $ n(k,x) $ in $ {\cal R}_2 $. We obtain especially 
in the regions where 
$l_{\rm sc}/L \sim \tau_{\rm ep}/\tau_{\rm op} $ that the neglection of the 
$ v_{\rm ph} $ term in (\ref{20}) is not allowed. Such parameter 
values lead to the result 
that the phonon distribution function at the boundary of the nanotube 
increases which is in fact seen in Fig.~5. In the regime 
$ {\cal R}_2 $ we obtain from (\ref{190}), (\ref{200})    
\begin{align} 
& {\cal R}_2 :  n_{{\cal R}_2} (k,x)=n_0(k,x_0(k))  
e^{K_1(x-x_0(k))} + 
\frac{s^p L}{v_{\rm ph} \tau_{\rm ep}} \frac{1}{K_1 L}  \nonumber \\
& \times 
\left[\frac{1}{K_1 L}\left(1+ K_1 x\right) - e^{K_1 (x- x_0(k))} 
\frac{1}{K_1 L} \left(1+ K_1 x_0(k)\right) \right] \label{220}
\end{align} 
where $ x_0(k)= L(1/2+ |1/2- v_F \hbar k/2 e U |)   $,  
$ n_0(k,x_0(k))= n_{{\cal R}_3} (k,x_0(k)) $ for 
$ v_F \hbar k/2 < e U/2 $  and $ n_0(k,x_0(k))=0 $ for 
$ v_F \hbar k/2 > e U/2 $. 

We now use the abbreviation 
\begin{equation} 
\tilde{K}_1=K_1 L \,.           \label{225} 
\end{equation} 
Next, we calculate the phonon density $ \overline{n}(x) $ (\ref{62}). 
According to the discussion above, we obtain 
\begin{align} 
&  \overline{n}(x)  \approx     
(\overline{n}_{{\cal R}_1}(L-x)+ \overline{n}_{{\cal R}_2}(x))  
\theta(\tilde{x}-1/2)  \label{230} \\
&  + (\overline{n}_{{\cal R}_1}(x) +\overline{n}_{{\cal R}_2}(L-x))  
\theta(1-(\tilde{x}-1/2))  
+2 \overline{n}_{{\cal R}_3}(x)   \nonumber   
 \end{align}  
where $ \overline{n}_{{\cal R}_i} $ are the energy averaged phonon 
distribution functions (\ref{210}), (\ref{215}) and (\ref{220}) according to (\ref{62}) where 
the integration region are restricted  to $ {\cal R}_i $ (\ref{180}). 
These are given by  
\begin{align} 
& \overline{n}_{{\cal R}_1}(x)= - \frac{1}{2} \frac{s^p L}{v_{\rm ph} \tau_{\rm ep}}
\frac{1}{\tilde{K}_1} \, \tilde{x}(1- 2\tilde{x}) ,   \label{240}  \\
& \overline{n}_{{\cal R}_3}(x)= -  
\frac{s^p L}{v_{\rm ph} \tau_{\rm ep}}
\frac{1}{\tilde{K}_1} \, \left(\frac{1}{2}-\left|\tilde{x}- \frac{1}{2} 
\right|\right)
\tilde{x}(1-\tilde{x})    \label{242}  
\end{align} 
and 
\begin{align} 
& \overline{n}_{{\cal R}_2}(x) \approx  
\frac{s^p L}{v_{\rm ph} \tau_{\rm ep}} 
\frac{1}{\tilde{K}_1} 
\left\{\left(\tilde{x}-\frac{1}{2}\right)
\left(\tilde{x}-1 \right) \right.
\label{250} \\
& \left.+\frac{1}{\tilde{K}_1} \! \left[ \! 
\left(2 \tilde{x}-\frac{3}{2} \right)
\! + \! \frac{1}{2} 
e^{\tilde{K}_1\left(\tilde{x}-\frac{1}{2}\right)} 
\right] + 
\frac{2}{\tilde{K}^2_1}\left(1-e^{\tilde{K}_1
\left(\tilde{x}-\frac{1}{2}\right)}\right) 
\right\} 
                     \nonumber 
\end{align} 
where we took into account that $ L/v_{\rm ph} \tau_{\rm op} \gg 1 $. 
Finally, we calculate the position averaged phonon distribution function 
$ \overline{n}= \int dx \, \overline{n}(x)/L $. We obtain 
\begin{equation} 
\overline{n} \approx 2 (\overline{n}_{{\cal R}_1}+ \overline{n}_{{\cal R}_2}
+ \overline{n}_{{\cal R}_3}) 
\label{255} 
\end{equation} 
with  
\begin{eqnarray}
 \overline{n}_{{\cal R}_1} & \approx & - 
\frac{s^p L}{v_{\rm ph} \tau_{\rm ep}} \frac{1}{\tilde{K}_1} \frac{1}{48} \,, 
\label{260}\\
 \overline{n}_{{\cal R}_3} & \approx & - 
\frac{s^p L}{v_{\rm ph} \tau_{\rm ep}} \frac{1}{\tilde{K}_1} \frac{13}{192}  
\label{262}
\end{eqnarray} 
and 
\begin{eqnarray} 
\overline{n}_{{\cal R}_2} & \approx & \frac{s^p L}{v_{\rm ph} \tau_{\rm ep}} 
\frac{1}{\tilde{K}_1}\left[- \frac{1}{48} + \frac{1}{ \tilde{K}^2_1}
\right.  \nonumber \\
 & & \left.  
+ \left(\frac{1}{2  \tilde{K}^2_1}- \frac{2}{ \tilde{K}^3_1}
\right) \left( e^{\tilde{K}_1/2}-1\right)\right] \,.  \label{265}
\end{eqnarray}
By taking into account the definition of $ \tilde{K}_1 $ (\ref{225}), 
(\ref{200}) we obtain 
that (\ref{255}) is the relation    
 which determines the energy-position averaged 
phonon distribution function $\overline{n} $ (\ref{65}). \\

Let us first assume that $ \tilde{K}_1 \gg 1 $ 
in region $ {\cal R}_2 $. From expression (\ref{265}) 
we find that there is no solution in this parameter region for (\ref{255}) 
since  $  L/v_{\rm ph} \tau_{\rm ep} \gg 1 $ and $ v_F \tau_{\rm ep} 
= 130 $nm.  

Next we assume that $ \tilde{K}_1 \ll -1 $ in the region 
$ {\cal R}_2 $. From (\ref{200}) we find then that 
$  \tilde{K}_1 \approx - L/v_{\rm ph} \tau_{\rm op} $ in the regions 
$ {\cal R}_i $. By using (\ref{255}) we obtain 
\begin{equation}    
\overline{n} \approx  \frac{13}{96} 
\frac{s^p \tau_{\rm op}}{\tau_{\rm ep}}  \quad \mbox{for} \quad 
L \gg \frac{v_{\rm ph} \tau_{\rm ep}}{\frac{26 s^p \tau_{\rm op}}
{96 \tau_{\rm ep}}+1}
\left(\frac{s^p \tau_{\rm op}}{\tau_{\rm ep}} -1 \right) .    
      \label{270}
\end{equation} 
For $ \tau_{\rm op} = 9.1$ps we obtain $ \overline{n} =5.3$.   
This leads to $ l_{\rm sc} = 11.1 $nm.
This value is in excellent agreement with the experimentally determined value 
 $ l_{\rm sc} \approx  10-11 $nm \cite{Javey1, Park1}. 
From (\ref{230}), (\ref{240}), (\ref{242})  and (\ref{250}) we obtain that 
the phonon distribution function 
$ \overline{n}(x) $ is zero at the boundary of the 
nanotube. By using the scattering parameters in (\ref{68}) we further obtain 
the validity of (\ref{270}) for $ L \gg 426 $nm. 

For smaller nanotube 
lengths one has $ \tilde{K}_1 \sim 0 $.       
By explicit calculation we obtain for $  L/v_{\rm ph} \tau_{\rm op} \gg 1 $
and $ \tilde{K}_1 \sim 0 $
that  
$ \overline{n}_{{\cal R}_1},  \overline{n}_{{\cal R}_3} 
 \ll \overline{n}_{{\cal R}_2} $. By taking only $ \overline{n}_{{\cal R}_2} $ into account in (\ref{230}) and (\ref{255}) we obtain 
for the energy averaged phonon distribution function by 
Taylor expanding (\ref{250})
with respect to $ \tilde{K}_1 $  
\begin{equation} 
\overline{n}(x) \approx  
\frac{s^p L}{v_{\rm ph} \tau_{\rm ep}}  
\left(\frac{1}{4} |\tilde{x}-1/2|^2 -\frac{1}{3}|\tilde{x}-1/2|^3 
\right)
\label{290}  
\end{equation} 
and for (\ref{255}) with (\ref{265}) 
\begin{equation} 
\overline{n}  \approx  
\frac{s^p L}{v_{\rm ph} \tau_{\rm ep}}  \frac{1}{96}   \,. 
\label{295}  
\end{equation}
From (\ref{290}) we get an increasing behavior of 
the phonon distribution function at 
the boundary of the nanotube in agreement with Fig.~5. \\

Finally, we calculate the phonon distribution function $ \overline{n} $ 
by solving (\ref{255}) for $ 
\overline{n} $ with (\ref{260}), (\ref{262})  and (\ref{265}) numerically as a 
function of the nanotube length $ L $. 
The result is shown in Fig.~7. We obtain that 
the decrease of $ \overline{n} $ for in the direction of small $ L $ 
given by (\ref{295}) is not very large such that we can assume 
approximately the validity of (\ref{270}) in the whole 
regime $ L \gtrsim 50 $nm. 
Next, we calculate the phonon distribution function 
for $ \pi eU \ll \hbar \omega L/l_{\rm sc} $.

\subsubsection{Phonon distribution function for   $ \pi eU \ll \hbar \omega 
L/l_{\rm sc} $}

By taking into account (\ref{110}), (\ref{150}) and (\ref{100}) we obtain 
that $ n(k,x) \not= n^{\rm ac}_B $ only in the regime 
$ |\pi (v_F |k| /2 \omega) l_{\rm sc}/L| \lesssim 1 $ where 
we use that $ \overline{n} \lesssim 1 $ as will be shown 
below. In this regime we obtain for $ k > 0 $ (\ref{190}) with 
 \begin{align}
& K_1= - \frac{1}{v_{\rm ph}}\left[\frac{s^p}{\tau_{\rm ep}} 
\frac{l_{\rm sc}}{L}\left(\frac{\cos(\pi \tilde{x})}{1-\cos^2(\pi \tilde{x})}
\right) + \frac{1}{\tau_{\rm op}}\right]  , 
\nonumber \\ 
 & K_2 = \frac{s^p}{4 v_{\rm ph}\tau_{\rm ep}}  
 \label{300} 
\end{align}
where again (\ref{178}) holds.        
We now assume that $ l_{\rm sc}/L \ll \tau_{\rm ep}/\tau_{\rm op} $ which is 
good fulfilled in the large length regime where we have (\ref{68}) 
$ \tau_{\rm ep}/\tau_{\rm op} \approx 0.22 $  . 
Then we obtain for 
$ |\pi (v_F |k| /2 \omega) l_{\rm sc}/L| \lesssim 1 $
\begin{equation} 
n(k,x) \approx  \frac{s^p}{4}
  \frac{\tau_{\rm op}}{\tau_{\rm ep}}\,. \label{310}
\end{equation}   
We point out that (\ref{310}) is not valid in the small regime 
where $ 0 \le |\epsilon| \le e U, \overline{n} \omega $. We obtain 
a position dependent numerical prefactor to (\ref{310}) in this regime. 
From (\ref{310}) we are able to calculate $ \overline{n}(x) $ and 
$ \overline{n} $ given by 
\begin{equation} 
\overline{n}(x)= \overline{n} \approx  \frac{s^p}{4} \frac{\tau_{\rm op}}{\tau_{\rm ep}}  \,. 
 \label{320}
\end{equation}

\end{appendix}



\begin{thebibliography}{99}
\bibitem{Rem1} 
We use a space and momentum grid with 500 points. The momentum range 
is then given by $ 5 e U/(\hbar v_F) $.
\end{thebibliography}
\end{document}